\documentclass{aa}  
\usepackage{color}
\usepackage{graphicx}
\usepackage{lscape}
\usepackage{natbib}
\usepackage[varg]{txfonts}
\usepackage{url}
\usepackage{xspace}
\bibpunct{(}{)}{;}{a}{}{,}
\def\etal{{et\,al.}\ }

\newcommand{\Teff}{$T\mathrm{\hspace*{-0.4ex}_{eff}}$\,}
\newcommand{\logg}{$\log\,g$\hspace*{0.5ex}}
\newcommand{\Msol}{M\hspace*{0.2ex}$_\odot$}

\newcommand{\lppr}{\stackrel{<}{\scriptstyle \sim}}
\newcommand{\lappr}{\raisebox{-0.4ex}{$\lppr $}}
\def\elf{PG\,1159$-$035}
\def\pgvier{PG\,1424+535}
\def\pgsieben{PG\,1707+427}

\def\re{RE\,0503$-$289}
\begin{document}

\title{The far-ultraviolet spectra of ``cool'' PG\,1159 stars}

\author{
K\@. Werner\inst{1} \and 
        T\@. Rauch\inst{1} \and
        J\@. W\@. Kruk\inst{2} 
}

\institute{Institute for Astronomy and Astrophysics, Kepler Center for Astro and
Particle Physics,  Eberhard Karls University, Sand~1, 72076
T\"ubingen, Germany\\ \email{werner@astro.uni-tuebingen.de}\and
           NASA Goddard Space Flight Center, Greenbelt, MD\,20771, USA}

\date{Received xx xx 2015 / Accepted xx xx 2015}

\authorrunning{K. Werner \etal}
\titlerunning{The far-ultraviolet spectra of ``cool'' PG\,1159 stars}

\abstract{We present a comprehensive study of Far Ultraviolet
  Spectroscopic Explorer (FUSE) spectra (912--1190\,\AA) of two
  members of the PG\,1159 spectral class, which consists of
  hydrogen-deficient (pre-) white dwarfs with effective temperatures
  in the range \Teff = 75\,000--200\,000\,K. As two representatives of
  the cooler objects, we have selected \pgsieben\ (\Teff = 85\,000\,K)
  and \pgvier\ (\Teff = 110\,000\,K), complementing a previous study
  of the hotter prototype \elf\ (\Teff = 140\,000\,K). The
  helium-dominated atmospheres are strongly enriched in carbon and
  oxygen, therefore, their spectra are dominated by lines from
  \ion{C}{iii-iv} and \ion{O}{iii-vi}, many of which were never observed
  before in hot stars. In addition, lines of many other metals (N, F,
  Ne, Si, P, S, Ar, Fe) are detectable, demonstrating that
  observations in this spectral region are most rewarding when
  compared to the near-ultraviolet and optical wavelength bands. We
  perform abundance analyses of these species and derive upper limits
  for several undetected light and heavy metals including iron-group
  and trans-iron elements. The results are compared to predictions of
  stellar evolution models for neutron-capture nucleosynthesis and
  good agreement is found.}


\maketitle
%

\section{Introduction}
\label{intro}

The overarching interest in PG\,1159 stars is related to the
investigation of nucleosynthesis processes in evolved low- to
intermediate-mass stars during their asymptotic giant branch (AGB)
stage. It is believed that these very hot, hydrogen-deficient (pre-)
white dwarfs (WDs) exhibit their helium-dominated layer as a
consequence of a late helium-shell flash \citep[e.g.,
][]{2006PASP..118..183W}. While usually kept hidden below the
hydrogen-rich envelope, these stars allow us to investigate directly
the chemistry of the H-He intershell region of the former red giant
star in which neutron-capture nucleosynthesis shapes the metal
abundance pattern. Spectroscopically, the access to metals heavier
than C, N, and O requires far-ultraviolet observations shortward of
about 1150\,\AA\ because the atoms are highly ionized. The search for
weak metal lines is particularly difficult for the PG\,1159 stars at
the ``cool'' end (effective temperature \Teff $\approx $ 75\,000 --
120\,000\,K) of the \Teff\ range they cover (up to 200\,000\,K). This
is because, as we  demonstrate in this paper,  a very large number of
spectral lines from the CNO elements in ionization stages
\ion{C}{iii}, \ion{N}{iv}, \ion{O}{iv-v}, appear in this lower
temperature interval while hotter objects, like the prototype
\elf\ \citep[140\,000\,K,][]{2007A&A...462..281J} exhibit
spectra with much fewer lines of the CNO elements appearing in higher
ionization stages (\ion{C}{iv}, \ion{N}{v}, and \ion{O}{vi}). Often,
heavier metals can only be detected by a handful of lines at best,
therefore, a good knowledge of potential line blends is important. 

Here we investigate two relatively cool PG\,1159 stars for which FUSE (Far
Ultraviolet Spectroscopic Explorer) spectra are available. We have two
aims. First we aim to compute model spectra   to identify as
many lines from the CNO elements as possible. This is initially
interesting in itself\ because the relatively high abundance of these
elements allows  the detection of many lines that are usually not seen
in stellar spectra. In addition, it poses a challenge because it
requires exceptionally large non-local thermodynamic equilibrium (non-LTE)
model atoms with atomic data to be gathered from different
sources. Second, based on these models we identify lines from heavier
metals and perform abundance determinations, taking 
potential line blends into account.

Preliminary results of the related, substantial work to achieve these
goals were published by our group in recent years as progress
reports
\citep{2005ASPC..334..225R,2007ASPC..372..237R,2008ASPC..391..121R,2013mueller}.
Based on this, we present here a comprehensive analysis with
considerably extended model atoms and hitherto not studied species,
leading to an unprecedented description and analysis of the
far-ultraviolet (far-UV) spectra of cool PG\,1159 stars. The paper is
organized as follows. We introduce the two program stars in
Sect.\,\ref{sect:programstars}. Their FUSE spectra and our method of
line identification are presented in
Sect.\,\ref{sect:observations}. Model atmosphere calculations for the
spectral analysis are specified in Sect.\,\ref{sect:models}. Details
of the line identification and results of our element abundance
analysis are described in Sect.\,\ref{sect:results}. A summary of the
results and a discussion in the context of abundance predictions from
stellar evolution theory in  Sect.\,\ref{sect:discussion} conclude the
paper.

\begin{figure*}[bth]
 \centering  \includegraphics[width=1.0\textwidth]{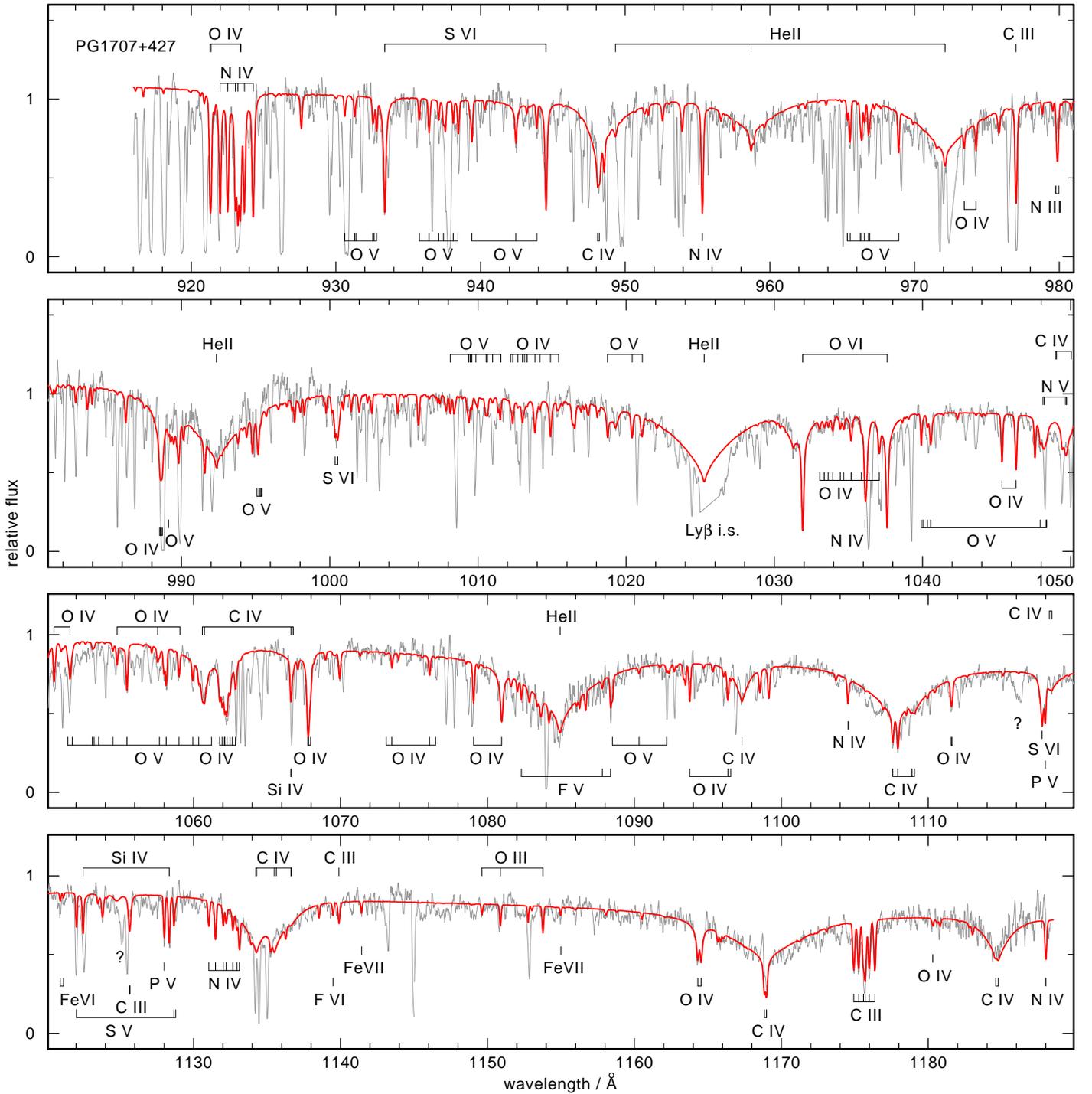}
  \caption{FUSE spectrum of \pgsieben\ { (thin line)} compared to a photospheric model spectrum  {(thick line)}
    with the finally adopted parameters as listed in
    Table\,\ref{tab:stars}. The main photospheric spectral lines are
    identified. Complete line lists are given in Tables\,\ref{tab:lines_c}--\ref{tab:lines}.
Question marks denote unidentified lines.}\label{fig:pg1707_fuse}
\end{figure*}

\begin{figure*}[bth]
 \centering  \includegraphics[width=1.0\textwidth]{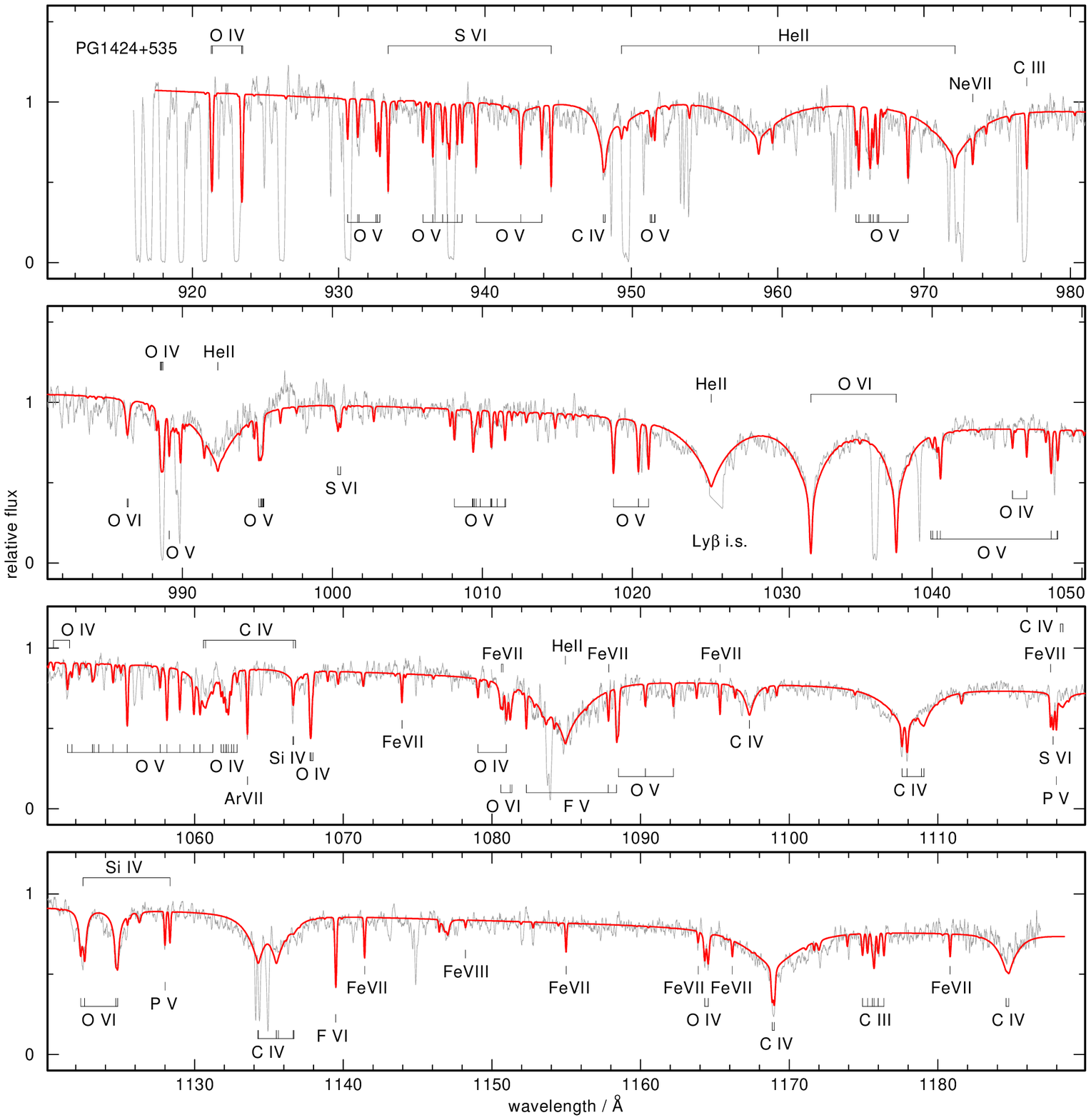}
  \caption{As in Fig.\,\ref{fig:pg1707_fuse}, for \pgvier.}\label{fig:pg1424_fuse}
\end{figure*}

\begin{table}
\begin{center}
\caption{Program stars and finally adopted
  parameters.\tablefootmark{a} }
\label{tab:stars} 
\tiny
\begin{tabular}{rrrr}
\hline 
\hline 
\noalign{\smallskip}
          & \pgsieben             &\pgvier               & Sun\tablefootmark{b}\\
\hline 
\noalign{\smallskip}
\Teff/\,K & 85\,000               & 110\,000 \\
\logg     & 7.5                   & 7.0\\
\noalign{\smallskip}
He        &  0.52                 & 0.52                 & 0.25\\
C         &  0.45                 & 0.45                 & $2.4 \times 10^{-3}$\\
N         &  0.01                 &$\leq3.5 \times 10^{-5}$&$6.9 \times 10^{-4}$\\
O         &  0.03                 & 0.03                 & $5.7 \times 10^{-3}$\\
F         &  $ 1.0 \times 10^{-4}$ & $ 5.0 \times 10^{-5}$ & $5.0 \times 10^{-7}$\\
Ne        &  \mbox{$-$\qquad}     & $ 1.0 \times 10^{-2}$ & $1.3 \times 10^{-3}$\\
Na        & \mbox{$-$\qquad}      & \mbox{$-$\qquad}     & $2.9 \times 10^{-5}$\\
Mg        &$\leq5.0 \times 10^{-3}$&$\leq5.0 \times 10^{-3}$&$7.1 \times 10^{-4}$\\
Al        &$\leq5.0 \times 10^{-3}$&$\leq5.0 \times 10^{-4}$&$5.6 \times 10^{-5}$\\
Si        &  $ 3.2 \times 10^{-4}$ & $ 2.0 \times 10^{-4}$ & $6.6 \times 10^{-4}$\\
P         &  $ 1.0 \times 10^{-5}$ & $ 3.2 \times 10^{-5}$ & $5.8 \times 10^{-6}$\\
S         &  $ 3.2 \times 10^{-4}$ & $ 1.0 \times 10^{-4}$ & $3.1 \times 10^{-4}$\\
Ar        &  \mbox{$-$\qquad}     & $ 6.0 \times 10^{-5}$ & $7.3 \times 10^{-5}$\\
Ca        &  \mbox{$-$\qquad}     & \mbox{$-$\qquad}     & $6.4 \times 10^{-5}$\\ 
\noalign{\smallskip}
Sc        &  \mbox{$-$\qquad}     & \mbox{$-$\qquad}     & $4.6 \times 10^{-8}$\\ 
Ti        &$\leq 3.1 \times 10^{-3}$& \mbox{$-$\qquad}    & $3.1 \times 10^{-6}$\\
V         &$\leq 3.2 \times 10^{-5}$&$\leq 3.2 \times 10^{-4}$& $3.2 \times 10^{-7}$\\
Cr        &$\leq 1.7 \times 10^{-5}$&$\leq 1.7 \times 10^{-4}$& $1.7 \times 10^{-5}$\\ 
Mn        &$\leq 1.1 \times 10^{-4}$&$\leq 1.1 \times 10^{-4}$& $1.1 \times 10^{-5}$\\ 
Fe        &$\leq 1.3 \times 10^{-3}$&$ 1.3 \times 10^{-3}$ & $1.3 \times 10^{-3}$\\
Co        &$\leq 4.2 \times 10^{-4}$&$\leq 4.2 \times 10^{-4}$& $4.2 \times 10^{-6}$\\
Ni        &$\leq 7.1 \times 10^{-5}$&$\leq 2.0 \times 10^{-4}$& $7.1 \times 10^{-5}$\\
\noalign{\smallskip}
Zn        &$\leq 5.0 \times 10^{-6}$&$\leq 5.0 \times 10^{-6}$& $1.7 \times 10^{-6}$\\
Ga        &$\leq 5.6 \times 10^{-5}$&$\leq 1.0 \times 10^{-5}$& $5.6 \times 10^{-8}$\\
Ge        &$\leq 2.4 \times 10^{-5}$&$\leq 1.0 \times 10^{-5}$& $2.4 \times 10^{-7}$\\
Kr        &$\leq 4.0 \times 10^{-4}$&$\leq 2.0 \times 10^{-5}$& $1.1 \times 10^{-7}$\\
Xe        &$\leq 1.0 \times 10^{-5}$&$\leq 2.0 \times 10^{-6}$& $1.7 \times 10^{-8}$\\
Ba        &$\leq 1.5 \times 10^{-3}$&$\leq 1.5 \times 10^{-3}$& $1.5 \times 10^{-8}$\\
\noalign{\smallskip} \hline
\end{tabular} 
\tablefoot{ 
  \tablefoottext{a}{Abundances in mass fractions and surface
    gravity $g$ in cm\,s$^{-2}$. If no abundance value is given, then
    no meaningful upper limit could be determined.}
  \tablefoottext{b}{Solar abundances from
    \citet{2009ARA&A..47..481A}.}  } 
\end{center}
\end{table}

\begin{table}[!ht]
\begin{center}
\caption{
Carbon lines identified in the FUSE spectra of at least one of our program stars.\tablefootmark{a}
  }
\label{tab:lines_c} 
\tiny
\begin{tabular}{llc}
\hline 
\hline 
\noalign{\smallskip}
Wavelength & Ion & Transition \\ \hline
\noalign{\smallskip} 
977.02  & \ion{C}{iii}  & $2{\rm s}^2\ ^1{\rm S} - 2{\rm s}2{\rm p}\ ^1{\rm P}^{\rm o}$ \\
1001.99:& \ion{C}{iii}  & $3{\rm p}\ ^1{\rm P}^{\rm o} - 6{\rm d}\ ^1{\rm D}$ \\
1016.34, 1016.40, 1016.53 & \ion{C}{iii}  & $3{\rm p}\ ^3{\rm P}^{\rm o}_{*} - 6{\rm d}\ ^3{\rm D}_{*}$ \\
1125.63--1125.68 & \ion{C}{iii}  & $3{\rm d}\ ^3{\rm D}_{*} - 6{\rm f}\ ^3{\rm F}^{\rm o}_{*}$ \\
1139.90 & \ion{C}{iii}  & $3{\rm p}\ ^1{\rm P}^{\rm o} - 5{\rm d}\ ^1{\rm D}$ \\
1165.62--1165.88\tablefootmark{b}& \ion{C}{iii}  & $3{\rm p}\ ^3{\rm P}^{\rm o} - 5{\rm d}\ ^3{\rm D}$\\
1174.93, 1175.26, 1175.59, \\ \quad 1175.71, 1175.99, 1176.37  & \ion{C}{iii}  & $2{\rm s}2{\rm p}\ ^3{\rm P}^{\rm o} - 2{\rm p}^2\ ^3{\rm P}$ \\
\noalign{\smallskip} 
 948.09, 948.21 & \ion{C}{iv}  &  $3{\rm s}- 4{\rm p}$ \\ 
1048.98\tablefootmark{d}, 1049.02\tablefootmark{d}  & \ion{C}{iv}  & $ 4{\rm d} - 11{\rm f}$ \\
1050.04         \tablefootmark{d}  & \ion{C}{iv}  & $ 4{\rm f} - 11{\rm g}$ \\
1060.59, 1060.74 &   \ion{C}{iv} &  $ 4{\rm p} - 10{\rm d}$ \\ 
1066.63, 1066.78\tablefootmark{c}&   \ion{C}{iv} & $ 4{\rm p} - 10{\rm s}$ \\ 
1097.32, 1097.34 &   \ion{C}{iv}  & $ 4{\rm s} - 8{\rm p}$ \\ 
1107.59, 1107.93 &   \ion{C}{iv}  & $ 3{\rm p}- 4{\rm d}$ \\ 
1108.89:, 1109.06: & \ion{C}{iv}  & $ 4{\rm p} - 9{\rm d}$ \\ 
1118.25:, 1118.41: & \ion{C}{iv}  & $ 4{\rm p} - 9{\rm s}$ \\ 
1134.25, 1134.30 &   \ion{C}{iv}  & $ 4{\rm d}- 9{\rm f}$ \\ 
1135.50 &   \ion{C}{iv}  &  $ 4{\rm f}- 9{\rm g}$ \\ 
1135.64 &   \ion{C}{iv}  &  $ 4{\rm f} - 9{\rm d}$ \\ 
1136.63, 1136.68 &   \ion{C}{iv}  & $ 4{\rm d} - 9{\rm p}$ \\ 
1168.85, 1168.99 &   \ion{C}{iv}  & $ 3{\rm d} - 4{\rm f}$ \\ 
1184.59, 1184.77 &   \ion{C}{iv}  & $ 4{\rm p} - 8{\rm d}$ \\ 
\noalign{\smallskip} \hline
\end{tabular} 
\tablefoot{
\tablefoottext{a}{Wavelengths in \AA. Colons denote uncertain identifications of very weak or blended lines}
\tablefoottext{b};{blend \ion{Fe}{vi}}
\tablefoottext{c}{and blend \ion{Si}{iv}}
\tablefoottext{d}{are not included in model spectra.}
} 
\end{center}
\end{table}

\begin{table}[!ht]
\begin{center}
\caption{
Nitrogen lines identified in \pgsieben.
  }
\label{tab:lines_n} 
\tiny
\begin{tabular}{llc}
\hline 
\hline 
\noalign{\smallskip}
Wavelength & Ion & Transition \\ \hline
\noalign{\smallskip} 
979.77, 979.83, 979.90, 979.97 & \ion{N}{iii} & $2{\rm s}2{\rm p}^2\ ^2{\rm D} - 2{\rm p}^3\,^2{\rm D}^{\rm o}$\\
989.79:\tablefootmark{a}, 991.51:\tablefootmark{b}, 991.58:\tablefootmark{b} & \ion{N}{iii} & $2{\rm s}^22{\rm p}\ ^2{\rm P}^{\rm o} - 2{\rm s}2{\rm p}^2\ ^2{\rm D}$\\
1005.99:, 1006.04:& \ion{N}{iii} & $2{\rm s}2{\rm p}^2\ ^2{\rm S} - 2{\rm p}^3\,^2{\rm P}^{\rm o}$\\
1183.03:& \ion{N}{iii} & $2{\rm s}2{\rm p}^2\ ^2{\rm P}_{1/2} - 2{\rm p}^3\,^2{\rm P}^{\rm o}_{1/2}$\\
\noalign{\smallskip} 
921.99, 922.52, 923.06, \\ \quad 923.22, 923.68, 924.28   & \ion{N}{iv}  & $2{\rm s}2{\rm p}\ ^3{\rm P}^{\rm o} - 2{\rm p}^2\ ^3{\rm P}$\\
948.15, 948.24, 948.29, \\ \quad 948.54, 948.56, 948.61 \tablefootmark{c} & \ion{N}{iv}  & $3{\rm p}\ ^3{\rm P}^{\rm o} - 4{\rm d}^2\ ^3{\rm D}$\\
955.33  & \ion{N}{iv}  & $2{\rm s}2{\rm p}\ ^1{\rm P}^{\rm o} - 2{\rm p}^2\ ^1{\rm S}$\\
1036.12--1036.33 & \ion{N}{iv}  & $3{\rm d}\ ^3{\rm D} - 4{\rm f}\ ^3{\rm F}^{\rm o}$\\
1078.71 & \ion{N}{iv}  & $3{\rm d}\ ^1{\rm D} - 4{\rm f}\ ^1{\rm F}^{\rm o}$\\
1104.54 & \ion{N}{iv}  & $3{\rm p}\ ^1{\rm P}^{\rm o} - 4{\rm s}\ ^1{\rm S}$\\
1117.93 & \ion{N}{iv}  & $2{\rm s}3{\rm d}\ ^1{\rm D} - 2{\rm p}3{\rm d}\ ^1{\rm P}^{\rm o}$\\
1123.81 & \ion{N}{iv}  & $3{\rm p}\ ^1{\rm D} - 4{\rm d}\ ^1{\rm D}^{\rm o}$\\
1131.04, 1131.49, 1132.02,  \\ \quad 1132.22, 1132.68, 1132.94 & \ion{N}{iv}  & $2{\rm s}3{\rm p}\ ^3{\rm P}^{\rm o} - 2{\rm p}3{\rm p}\ ^3{\rm P}$\\
1133.12, 1135.25, 1136.27 & \ion{N}{iv}  & $2{\rm s}3{\rm s}\ ^3{\rm S} - 2{\rm p}3{\rm s}\ ^3{\rm P}^{\rm o}$\\
1188.00: & \ion{N}{iv}  & $2{\rm s}3{\rm s}\ ^1{\rm S} - 2{\rm p}3{\rm s}\ ^1{\rm P}^{\rm o}$\\
\noalign{\smallskip} 
1018.97, 1019.29, 1019.31 & \ion{N}{v} & $4{\rm p} - 6{\rm d}$ \\
1048.13, 1048.23          & \ion{N}{v} & $4{\rm d} - 6{\rm f}$ \\
1049.65, 1049.71          & \ion{N}{v} & $4{\rm f} - 6{\rm g}$ \\
\noalign{\smallskip} \hline
\end{tabular} 
\tablefoot{
\tablefoottext{a}{Dominated by \ion{N}{iii} i.s.}
\tablefoottext{b};{dominated by H$_2$ i.s}
\tablefoottext{c};{blend with \ion{C}{iv.}}
} 
\end{center}
\end{table}

\begin{table}[!ht]
\begin{center}
\caption{Oxygen lines identified in at least one of our program stars.}
\label{tab:lines_o} 
\tiny
\begin{tabular}{llc}
\hline 
\hline 
\noalign{\smallskip}
Wavelength & Ion & Transition \\ \hline
\noalign{\smallskip} 
1149.63, 1150.88, 1153.77 & \ion{O}{iii} &  $2{\rm s}2{\rm p}^3\,^3{\rm S}^{\rm o} - 2{\rm p}^4\,^3{\rm P}$\\ 
\noalign{\smallskip} 
 921.30, 921.37, 923.37, 923.44 & \ion{O}{iv}  & $2{\rm s}2{\rm p}^2\,^2{\rm P} - 2{\rm p}^3\,^2{\rm P}^{\rm o}$\\ 
 938.47 & \ion{O}{iv}  & $3{\rm p}\,^2{\rm P}_{3/2} - 4{\rm s}\,^2{\rm P}^{\rm o}_{3/2}$ \\
 973.43:\tablefootmark{a}, 974.25 & \ion{O}{iv}  & $2{\rm s}^2 3{\rm p}\,^2{\rm P}^{\rm o} - 2{\rm s}2{\rm p}3{\rm p}\,^2{\rm S}$\\ 
 994.80, 995.16, 995.17 & \ion{O}{iv} & $3{\rm p}\,^4{\rm D}_{*} - 4{\rm s}\,^4{\rm P}^{\rm o}_{*}$ \\
1013.01, 1013.86, 1014.90 & \ion{O}{iv}  & $3{\rm d}\,^4{\rm F}^{\rm o}_{*} - 4{\rm f}\,^4{\rm F}_{*}$ \\ 
1035.18, 1037.09 & \ion{O}{iv}  & $3{\rm d}\,^4{\rm D}^{\rm o}_{*} - 4{\rm f}\,^4{\rm D}_{*}$ \\ 
1045.36, 1046.31 & \ion{O}{iv}  & $2{\rm s}^2 3{\rm p}\,^2{\rm P}^{\rm o} - 2{\rm s}^2 4{\rm s}\,^2{\rm S}$\\ 
1047.59, 1050.50& \ion{O}{iv}  & $2{\rm s}^2 3{\rm s}\,^2{\rm S} - 2{\rm s}2{\rm p}3{\rm s}\,^2{\rm P}^{\rm o}$\\ 
1051.59 & \ion{O}{iv}  & $3{\rm d}\,^2{\rm D}^{\rm o}_{5/2} - 4{\rm f}\,^2{\rm D}_{5/2}$ \\
1054.80, 1057.55, 1059.07  & \ion{O}{iv} & $3{\rm p}\,^4{\rm S} - 4{\rm s}\,^4{\rm P}^{\rm o}$ \\
1061.78, 1061.95, 1062.10, \\ \quad 1062.13, 1062.27,  \\ \quad 1062.48, 1062.86 & \ion{O}{iv}  & $3{\rm d}\,^4{\rm D}^{\rm o} - 4{\rm f}\,^4{\rm F}$ \\
1067.77, 1067.83, 1067.96& \ion{O}{iv}  & $3{\rm d}\,^2{\rm D} - 4{\rm f}\,^2{\rm F}^{\rm o}$ \\
1079.06, 1081.02, 1082.04:& \ion{O}{iv}  & $2{\rm s}^2 3{\rm p}\,^2{\rm P}^{\rm o} - 2{\rm s}2{\rm p}3{\rm p}\,^2{\rm D}$\\ 
1080.97 & \ion{O}{iv}  & $3{\rm d}\,^2{\rm D}^{\rm o}_{*} - 4{\rm f}\,^2{\rm F}_{*}$ \\ 
1093.77, 1096.36, 1096.56:& \ion{O}{iv}  & $2{\rm s}^2 3{\rm d}\,^2{\rm D} - 2{\rm s}2{\rm p}3{\rm d}\,^2{\rm F}^{\rm o}$\\ 
1099.15:& \ion{O}{iv}  & $3{\rm p}\,^2{\rm D}_{5/2} - 4{\rm s}\,^2{\rm P}^{\rm o}_{3/2}$ \\
1111.56, 1111.63 & \ion{O}{iv} & $3{\rm p}\,^4{\rm P}_{*} - 4{\rm s}\,^4{\rm P}^{\rm o}_{*}$ \\
1164.32, 1164.55 & \ion{O}{iv} & $3{\rm d}\,^2{\rm F}^{\rm o}_{*} - 4{\rm f}\,^2{\rm G}_{*}$ \\ 
1180.31 & \ion{O}{iv}  & $2{\rm s}^2 4{\rm p}\,^2{\rm P}^{\rm o}_{3/2} - 2{\rm s}2{\rm p} 4{\rm p}\,^2{\rm D}_{5/2}$\\ 
\noalign{\smallskip} 
 931.30, 931.39, 932.64, 932.82 & \ion{O}{v}   & $2{\rm s}3{\rm d}\,^3{\rm D}_{*} - 2{\rm p}3{\rm d}\,^3{\rm P}^{\rm o}_{*}$\\ 
 935.76, 936.43, 937.11, 938.11,\\ \quad  938.43  & \ion{O}{v}   & $2{\rm s}3{\rm p}\,^3{\rm P}^{\rm o}_{*} - 2{\rm p}3{\rm p}\,^3{\rm P}_{*}$\\ 
 939.40, 942.44, 943.88  & \ion{O}{v}   & $2{\rm s}3{\rm s}\,^3{\rm S} - 2{\rm p}3{\rm s}\,^3{\rm P}^{\rm o}$\\
 951.28, 951.38, 951.59, 951.62  & \ion{O}{v}   & $4{\rm p}\,^3{\rm P}^{\rm o} - 6{\rm d}\,^3{\rm D}$ \\
 959.64 & \ion{O}{v}   & $4{\rm p}\,^1{\rm P}^{\rm o} - 6{\rm d}\,^1{\rm D}$ \\
 965.35, 965.54, 966.25, 966.34, \\ \quad 966.53, 966.81, 966.90 & \ion{O}{v}   & $2{\rm s}3{\rm d}\,^3{\rm D} - 2{\rm p}3{\rm d}\,^3{\rm D}^{\rm o}$\\
 968.90 & \ion{O}{v}   & $2{\rm s}3{\rm s}\,^1{\rm S} - 2{\rm p}3{\rm s}\,^1{\rm P}^{\rm o}$\\ 
 988.29:& \ion{O}{v}   & $2{\rm s}4{\rm d}\,^1{\rm D} - 2{\rm p}4{\rm d}\,^1{\rm P}^{\rm o}$\\ 
 989.12 & \ion{O}{v}   & $2{\rm s}3{\rm p}\,^3{\rm P}^{\rm o}_{1} - 2{\rm p}3{\rm p}\,^3{\rm S}_{1}$\\ 
 991.46 & \ion{O}{v}   & $2{\rm s}4{\rm d}\,^1{\rm D} - 2{\rm p}4{\rm d}\,^1{\rm F}^{\rm o}$\\ 
 995.09, 995.23, 995.34 & \ion{O}{v}   & $4{\rm d}\,^3{\rm D}_{*} - 6{\rm f}\,^3{\rm F}^{\rm o}_{*}$ \\
 996.53:& \ion{O}{v}   & $2{\rm s}3{\rm d}\,^1{\rm D} - 2{\rm p}3{\rm d}\,^1{\rm F}^{\rm o}$\\ 
1007.86, 1008.14, 1009.49,  \\ \quad 1009.59, 1009.87, 1010.99 & \ion{O}{v}   & $2{\rm s}4{\rm p}\,^3{\rm P}^{\rm o} - 2{\rm p}4{\rm p}\,^3{\rm P}$\\
1018.78, 1020.45, 1021.13 & \ion{O}{v}   & $2{\rm s}4{\rm f}\,^3{\rm F}^{\rm o} - 2{\rm p}4{\rm f}\,^3{\rm G}$\\
1040.30 & \ion{O}{v}   & $4{\rm d}\,^3{\rm P}_{3} - 6{\rm p}\,^3{\rm S}^{\rm o}_{*}$\\
1040.33 & \ion{O}{v}   & $2{\rm s}4{\rm p}\,^3{\rm P}^{\rm o}_{2} - 2{\rm p}4{\rm p}\,^3{\rm S}_{1}$\\
1040.56 & \ion{O}{v}   & $2{\rm s}4{\rm f}\,^1{\rm F}^{\rm o} - 2{\rm p}4{\rm f}\,^1{\rm G}$\\
1047.94 & \ion{O}{v}   & $2{\rm p}3{\rm d}\,^1{\rm F}^{\rm o} - 2{\rm s}5{\rm g}\,^1{\rm G}$\\
1048.38, 1051.45, 1051.75,  \\ \quad 1053.13, 1053.24 & \ion{O}{v}   & $2{\rm s}4{\rm p}\,^3{\rm P}^{\rm o}_{*} - 2{\rm p}4{\rm p}\,^3{\rm D}_{*}$\\
1055.47, 1058.14, 1059.01,  \\ \quad 1059.95, 1060.36 & \ion{O}{v}   & $2{\rm s}3{\rm p}\,^3{\rm P}^{\rm o}_{*} - 2{\rm p}3{\rm p}\,^3{\rm D}_{*}$\\ 
1071.34 & \ion{O}{v}   & $4{\rm d}\,^1{\rm D} - 6{\rm p}\,^1{\rm P}^{\rm o}$\\ 
1088.51\tablefootmark{b} & \ion{O}{v}   & $2{\rm s}3{\rm p}\,^1{\rm P}^{\rm o} - 2{\rm p}3{\rm p}\,^1{\rm P}$\\
1090.32 & \ion{O}{v}   & $2{\rm s}4{\rm p}\,^1{\rm P}^{\rm o} - 2{\rm p}4{\rm p}\,^1{\rm P}$\\
1092.20 & \ion{O}{v}   & $2{\rm s}4{\rm d}\,^1{\rm D} - 2{\rm p}4{\rm d}\,^1{\rm D}^{\rm o}$\\
\noalign{\smallskip} 
 986.32 & \ion{O}{vi}  & $4{\rm s}- 5{\rm p}$ \\
1031.91, 1037.61 & \ion{O}{vi}  & $2{\rm s}- 2{\rm p}$ \\  
1080.60, 1081.24, 1081.34  & \ion{O}{vi}  & $4{\rm p} - 5{\rm d}$ \\  
1122.34, 1122.60 & \ion{O}{vi}  & $4{\rm d} - 5{\rm f}$ \\
1124.70, 1124.82 & \ion{O}{vi}  & $4{\rm f} - 5{\rm g}$ \\  
1126.17, 1126.28& \ion{O}{vi}  & $4{\rm f} - 5{\rm d}$ \\  
1146.75, 1146.83, 1147.03 & \ion{O}{vi}  & $4{\rm d} - 5{\rm p}$ \\
1171.12, 1172.00 & \ion{O}{vi}  & $4{\rm p}- 5{\rm s}$ \\  
\noalign{\smallskip} \hline
\end{tabular} 
\tablefoot{
\tablefoottext{a}{blend with \ion{Ne}{vii}}
\tablefoottext{b};{blend with \ion{F}{v}}.
} 
\end{center}
\end{table}

\begin{table}[!ht]
\begin{center}
\caption{
Lines of F, Ne, Si, P, S, Ar, Fe, and Ni identified in at least one of our program stars.
  }
\label{tab:lines} 
\tiny
\begin{tabular}{llc}
\hline 
\hline 
\noalign{\smallskip}
Wavelength & Ion & Transition \\ \hline
\noalign{\smallskip} 
1082.31, 1087.82, 1088.39\tablefootmark{i} & \ion{F}{v}   & $2{\rm s}2{\rm p}^2\ ^2{\rm P} - 2{\rm p}^3\ ^2{\rm D}^{\rm o}$\\
1139.50 & \ion{F}{vi}  & $2{\rm s}2{\rm p}\ ^1{\rm P}^{\rm o} - 2{\rm p}^2\ ^1{\rm D}$ \\
\noalign{\smallskip} 
973.33\tablefootmark{h}  & \ion{Ne}{vii}& $2{\rm s}2{\rm p}\ ^1{\rm P}^{\rm o} - 2{\rm p}^2\ ^1{\rm D}$ \\
\noalign{\smallskip} 
1066.61--1066.65\tablefootmark{k} & \ion{Si}{iv} & $3{\rm d}\ ^2{\rm D} - 4{\rm f}\ ^2{\rm F}^{\rm o}$ \\ 
1122.49\tablefootmark{l}, 1128.34 & \ion{Si}{iv} & $3{\rm p}\ ^2{\rm P}^{\rm o} - 3{\rm d}\ ^2{\rm D}$ \\ 
1118.81: & \ion{Si}{v}  & $3{\rm s}\ ^3{\rm P}^{\rm o}_{2}-3{\rm p}\ ^3{\rm P}_{2}$\\ 
\noalign{\smallskip} 
1000.38:\tablefootmark{a} & \ion{P}{v}   & $3{\rm d}\ ^2{\rm D}_{3/2}-4{\rm p}\ ^2{\rm P}_{1/2}^{\rm o}$ \\
1117.98\tablefootmark{g}, 1128.01\tablefootmark{m} & \ion{P}{v}   & $3{\rm s}\ ^2{\rm S}-3{\rm p}\ ^2{\rm P}^{\rm o}$ \\
\noalign{\smallskip} 
1039.92\tablefootmark{b}& \ion{S}{v}   & $3{\rm s}3{\rm d}\ ^1{\rm D} - 3{\rm p}3{\rm d}\ ^1{\rm F}^{\rm o}$ \\ 
1122.03:\tablefootmark{d}, 1128.67,  \\ \quad 1128.78, 1133.90:  & \ion{S}{v}   & $3{\rm s}3{\rm d}\ ^3{\rm D} - 3{\rm p}3{\rm d}\ ^3{\rm F}^{\rm o}$ \\
933.38, 944.52  & \ion{S}{vi}  & $3{\rm s} - 3{\rm p} $ \\
1000.37, 1000.54  & \ion{S}{vi}  & $4{\rm d} - 5{\rm f}$ \\
1117.76 & \ion{S}{vi}  & $4{\rm f} - 5{\rm g} $ \\
\noalign{\smallskip} 
1063.55\tablefootmark{n} & \ion{Ar}{vii}& $3{\rm s}3{\rm p}\ ^1{\rm P}^{\rm o} - 3{\rm p}^2\ ^1{\rm D}$ \\
\noalign{\smallskip} 
1005.96:\tablefootmark{p} & \ion{Fe}{v}  & $4{\rm p}\ ^5{\rm F}^{\rm o}_{11} - 4{\rm d}\ ^5{\rm G}_{13}$\\
1115.10: & \ion{Fe}{vi} & $4{\rm s}\ ^2{\rm D}_{5/2} - 4{\rm p}\ ^2{\rm D}_{5/2}^{\rm o}$\\
1120.93:, 1121.15:  & \ion{Fe}{vi}  & $4{\rm s}\ ^2{\rm F} - 4{\rm p}\ ^2{\rm F}^{\rm o}$\\
1152.77:\tablefootmark{e} & \ion{Fe}{vi}  & $4{\rm s}\ ^2{\rm G}_{7/2} - 4{\rm p}\ ^2{\rm F}_{5/2}^{\rm o}$\\
1160.51:, 1165.67:\tablefootmark{f} & \ion{Fe}{vi}  & $4{\rm s}\ ^2{\rm P} - 4{\rm p}\ ^2{\rm P}^{\rm o}$ \\
1073.95 & \ion{Fe}{vii} & $4{\rm s}\ ^1{\rm D}-4{\rm p}\ ^1{\rm P}^{\rm o}      $ \\
1080.64\tablefootmark{o}, 1080.74\tablefootmark{o}, \\ \quad 1087.86, 1095.34 & \ion{Fe}{vii}& $4{\rm s}\ ^3{\rm D}-4{\rm p}\ ^3{\rm P}^{\rm o}  $ \\ 
1117.58 & \ion{Fe}{vii}& $4{\rm s}\ ^1{\rm D}-4{\rm p}\ ^1{\rm F}^{\rm o}      $ \\ 
1141.43, 1154.99, 1166.17: & \ion{Fe}{vii}& $4{\rm s}\ ^3{\rm D}-4{\rm p}\ ^3{\rm F}^{\rm o}  $ \\
1163.88, 1180.83:& \ion{Fe}{vii} & $4{\rm s}\ ^3{\rm D}-4{\rm p}\ ^3{\rm D}^{\rm o}  $ \\
1062.44:\tablefootmark{c}& \ion{Fe}{viii}& $4{\rm s}\ ^2{\rm S}_{1/2}-4{\rm p}\ \ ^2{\rm P}^{\rm o}_{3/2}$ \\
1148.22:& \ion{Fe}{viii}& $4{\rm s}\ ^2{\rm S}_{1/2}-3{\rm d}^2\ ^2{\rm P}^{\rm o}_{3/2}$ \\
\noalign{\smallskip} 
1075.80:& \ion{Ni}{vi}  & $4{\rm s}\ ^2{\rm I}_{11/2} - 4{\rm p}\ ^2{\rm H}_{9/2}^{\rm o}$\\
1124.19:& \ion{Ni}{vi}  & $4{\rm s}\ ^2{\rm H}_{13/2} - 4{\rm p}\ ^2{\rm I}_{15/2}^{\rm o}$\\  
1144.03:& \ion{Ni}{vi}  & $4{\rm s}\ ^2{\rm H}_{9/2} - 4{\rm p}\ ^2{\rm I}_{11/2}^{\rm o}$\\  
1154.60:& \ion{Ni}{vi}  & $4{\rm s}\ ^6{\rm D}_{7/2} - 4{\rm p}\ ^6{\rm F}_{9/2}^{\rm o}$\\  
\noalign{\smallskip} \hline
\end{tabular} 
\tablefoot{
\tablefoottext{a}{blend \ion{S}{vi}}
\tablefoottext{b};{blend \ion{O}{v}}
\tablefoottext{c};{blend \ion{O}{iv}}
\tablefoottext{d};{blend \ion{Fe}{ii} i.s.}
\tablefoottext{e};{blend \ion{C}{ii} i.s.}
\tablefoottext{f};{blend \ion{C}{iii}}
\tablefoottext{g};{blend \ion{N}{iv}}
\tablefoottext{h};{blend \ion{O}{iv}}
\tablefoottext{i};{blend \ion{O}{v}}
\tablefoottext{k};{blend \ion{Ar}{i} i.s. and \ion{C}{iv}}
\tablefoottext{l};{blend \ion{C}{i} i.s.}
\tablefoottext{m};{blend \ion{Fe}{ii} i.s.}
\tablefoottext{n};{blend H$_2$ i.s. in \pgsieben}
\tablefoottext{o};{blend \ion{O}{iv} and \ion{O}{vi}}
\tablefoottext{p}; and {blend \ion{N}{iii}}.
} 
\end{center}
\end{table}

\begin{table}
\begin{center}
\caption{Number of levels and lines of model ions used for line formation calculations of metals.\tablefootmark{a} }
\label{tab:modelatoms} 
\tiny
\begin{tabular}{cccccccc}
\hline 
\hline 
\noalign{\smallskip}
   &  II   &  III      &    IV    &    V     &     VI  &    VII \\
\hline 
\noalign{\smallskip}
C  &   1,0 &   133,745 &   54,279 &   \\   
N  &       &   34,129  &   76,405 &   54,297 & \\
O  &       &   47,171  &   83,637 &   105,671&   9,15  \\
F  &       &           &   1,0    &   15,31  &   15,27 &    2,1   \\
Ne &       &           &          &   14,18  &   14,30 &    15,27 \\
Na &       &           &   49,39  &   52,229 &   65,301&    29,80 \\  
Mg &       &   1,0     &   31,93  &   52,175 &   26,55 &    46,147\\
Al &       &   1,0     &   15,20  &   30,87  &   43,126&    43,160\\
Si &       &   17,28   &   30,102 &   25,59  \\
P  &       &   3,0     &   21,9   &   18,12  \\
S  &       &           &   17,32  &   39,107 &   25,48 \\
Ar &       &           &          &   4,0    &   48,225&    40,130\\
Kr &       &           &   38,0   &  25,0    &   46,887&    14,37 \\
Xe &       &           &   44,0   &  29,0    &   82,887&    36,124\\
\noalign{\smallskip} \hline
\end{tabular} 
\tablefoot{ \tablefoottext{a}{First and second number of each table
    entry denote the number of levels and lines, respectively. The the highest ionization stage, which only comprises its ground state, is not listed
    for each element.
    For the treatment of iron-group and trans-iron elements (except of Kr and Xe), see text.}  } 
\end{center}
\end{table}

\section{The program stars}
\label{sect:programstars}

\pgsieben\ and \pgvier\ were among the first stars classified as
PG\,1159 stars
\citep{1980PASP...92..657L,1984NASCP2349..273S,1985ApJS...58..379W}
and constituted two of the four PG\,1159 stars that were first
analysed with non-LTE model atmospheres
\citep{1991A&A...244..437W}. Based on optical spectra, identical
values for the basic atmospheric parameters (\Teff\ and surface
gravity $g$) were derived for both objects: \Teff = 100\,000\,K, \logg
= 7 ($g$ in cm\,s$^{-2}$). Only a coarse estimate on the O abundance was
possible, based on the \ion{O}{v}~1371\,\AA\ line in poor
signal-to-noise ratio (S/N) low-resolution  International
Ultraviolet Explorer (IUE) spectra. He/C/O abundances (given in mass
fractions throughout the paper) of 0.33/0.50/0.17 were
derived. Medium-resolution UV spectroscopy performed with the Goddard High Resolution Spectrograph 
(GHRS) aboard the Hubble Space
Telescope (HST) enabled \citet{1998A&A...334..618D} to refine these
parameters. In particular, they found that \pgsieben\ must be
significantly cooler than \pgvier. For \pgsieben\ \Teff/\logg =
85\,000/7.5 was derived and 110\,000/7.0 for \pgvier. 

As to the nitrogen abundance, a dichotomy was found by
\citet{1998A&A...334..618D}. From their UV spectral analysis of nine
PG\,1159 stars they found that four typically have  N = 0.035, while
the others have at least two or three orders of magnitude less
nitrogen. \pgsieben\ and \pgvier\ were found to belong to the former
and latter groups, respectively. The abundance determination of other
trace elements was only possible with the advent of FUSE. Spectral
lines of F, Ne, Si, P, S, Ar, and Fe were identified, as we
discuss in detail below (Sect.\,\ref{sect:results}).

\pgsieben\ and \pgvier\ are the only PG\,1159 stars with
\Teff\ $\lappr$ 110\,000\,K that were observed with FUSE. Two other
objects of this spectral class with \Teff = 110\,000\,K, the central
stars of the planetary nebulae Abell~43 and NGC\,7094, were also
observed, however, their surface gravity is much lower (hence their
luminosity much higher) so that the ionization balances of all metals
are shifted significantly to higher ionization stages and their line
spectra are qualitatively different, including the occurrence of
P~Cygni line profiles \citep{2010AIPC.1273..231F}.

\section{Observations and line identifications}
\label{sect:observations}

The raw FUSE data for the  program stars were retrieved from the
Mikulski Archive for Space Telescopes (MAST). The observation IDs were
P1320301 for \pgvier\ and P1320401 for \pgsieben. Both were observed
through the LWRS spectrograph aperture, and the data were obtained in
time-tag mode.

The SiC1 channel was slightly misaligned for the observation of
\pgvier, as the SiC1 fluxes varied $\approx 15$\,\% from exposure to
exposure, and the net flux was $\approx 30$\,\% below that measured in
the other channels. The spectrum in this channel was renormalized to
match that of the LiF1 spectrum at wavelengths where the spectral
coverage overlapped. In all other cases, the fluxes measured in any
given channel were consistent to 1\% or better from exposure to
exposure, indicating good alignment, and consistent to within a few
percent when comparing one channel to another. The net exposure times
were 10.4\,ksec for \pgvier\ and 14.3\,ksec for \pgsieben.

Raw data were processed twice with CalFUSE v3.2.3: once with screening
parameters set to extract data only during orbital Night, and once to
extract data during orbital Day. Zero-point offsets in the wavelength
scale were adjusted for each exposure by shifting each spectrum to
coalign narrow interstellar absorption features. The individual
exposures from each observation were then combined to form composite
Day and Night spectra for each channel. The Day and Night spectra were
then compared at the locations of all the known airglow emission
lines. If the Day spectra showed any excess flux in comparison to the
Night spectra at those wavelengths, the corresponding pixels in the
Day spectra were flagged as bad and were not included in subsequent
processing. Significant airglow was present during orbital Night only
at Lyman $\beta$ and at Lyman $\gamma$; faint emission can be
discerned at Lyman $\delta$. This residual emission would affect fits
to the interstellar absorption at these wavelengths, but has no impact
on any of the analyses we present.

The final step was to combine the spectra from the four instrument
channels into a single composite spectrum. Because of residual
distortions in the wavelength scale in each channel, additional shifts
of localized regions of each spectrum were required to coalign the
spectra; these shifts were typically only 1--2 pixels. Bad pixels
resulting from detector defects were flagged at this point and
excluded from further processing. Finally, the spectra were resampled
onto a common wavelength scale and combined, weighting by S/N on a
pixel-by-pixel basis.

The final spectrum has S/N = 11--23 per 0.013A pixel for \pgvier, and
9--15 per pixel for \pgsieben. The effects of fixed-pattern noise are
minimized by the fact that the positions of the spectra on the
detectors varied during each observation, and by the fact that nearly
every wavelength bin was sampled by at least two different detectors.

The first aim of our study is to identify, as far as possible, the
photospheric spectral lines in the FUSE spectra of \pgsieben\ and
\pgvier\ (Figs.\,\ref{fig:pg1707_fuse} and \ref{fig:pg1424_fuse}). A
similar study was performed by \citet{2007A&A...462..281J} for the
prototype of the PG\,1159 class, \elf, which is significantly hotter
(\Teff = 140\,000\,K). They found that carbon and oxygen are exclusively
detected by \ion{C}{iv} and \ion{O}{vi} lines. In contrast, as we
shall see, the spectra of our cooler program stars are dominated by
lower ionization stages, namely, \ion{C}{iii} and \ion{O}{iii-v}.

Our procedure for line identification is strongly based on our
synthetic model atmosphere spectra. Since the basic atmospheric
parameters (\Teff\ and \logg) are well known, the main task in the
analysis of C and O is to employ large model atoms, i.e., many energy
levels and radiative transitions to predict the line spectra
as completely as possible. In addition, for \pgsieben, the N spectrum
can be studied in detail. This leads to a surprisingly large number of
identified lines, in comparison to the hot PG\,1159 stars, and most of
them were never seen in hot stars. The other metals show many fewer
lines because they are just trace elements. Only their
strongest lines are detectable, which usually are transitions between
relatively low excited levels so that smaller model atoms suffice. In
Tables~\ref{tab:lines_c}--\ref{tab:lines_o}, we present all
photospheric lines of C, N, and O, which we identified in at least one
of our two program stars. Table\,\ref{tab:lines} summarizes the
lines from other metals; some of them were detected in earlier
work. They are assessed in detail below (Sect.\,\ref{sect:results}).

In the spectra of both stars, some likely photospheric lines remain
unidentified. Two rather prominent examples in \pgsieben\ are located
at 1116.2 and 1125.0\,\AA. They might stem from \ion{N}{iv} multiplets
with unknown wavelength position. These kinds of multiplets are predicted by our
models, which include (computed) OP energy level data but  do not
appear in any list of observed line positions. We treat them like
Kurucz's LIN lines, namely, we do not include them in our final
synthetic spectra.

To identify lines from the interstellar medium (ISM) and to judge
their potential contamination of photospheric lines, we used the
program OWENS \citep{2002P&SS...50.1169H,2003ApJ...599..297H} which
considers different clouds with individual radial and turbulent
velocities, temperatures, column densities, and chemical
compositions. Note that we did not aim to make good fits to the ISM
lines.  For \pgsieben, our model includes lines of \ion{D}{i},
\ion{H}{i}, H$_2$, \ion{C}{i--iii}, \ion{N}{i-iii}, \ion{O}{i},
\ion{P}{ii}, \ion{S}{ii--iii}, \ion{Ar}{i}, and \ion{Fe}{ii}. For
\pgvier, our model includes lines of \ion{D}{i}, \ion{H}{i}, H$_2$
(J=0,1), \ion{C}{i-iii}, \ion{N}{i-iii}, \ion{O}{i}, \ion{Si}{ii},
\ion{P}{ii}, \ion{Ar}{i}, and \ion{Fe}{ii} in two clouds with radial
velocities of $-8$\,km/s and $-66$\,km/s. This line of sight is
unusual in that H$_2$ was just barely detectable. Figures
\ref{fig:pg1707_fuse_ism} and \ref{fig:pg1424_fuse_ism} in the
Appendix are identical to Figs.\,\ref{fig:pg1707_fuse} and
\ref{fig:pg1424_fuse}, but include the ISM lines. 

\section{Model atoms and model atmospheres}
\label{sect:models}

For the spectral analysis, we used our non-LTE code to compute
plane-parallel, line-blanketed atmosphere models in radiative and
hydrostatic equilibrium
\citep{1999JCoAM.109...65W,2003ASPC..288...31W}. They include the
three most abundant elements, namely He, C, and O. All other species
were treated one by one as trace elements, i.e., keeping fixed the
atmospheric structure. In the same manner, extended model atoms for C
and O were introduced, meaning that non-LTE population numbers were
computed for highly-excited levels, which were treated in LTE during the
preceding model-atmosphere computations. Table\,\ref{tab:modelatoms}
summarizes the number of considered non-LTE levels and radiative
transitions between them. All model atoms were built from the publicly
available T\"ubingen Model Atom Database (TMAD\footnote{\tt
  http://astro.uni-tuebingen.de/{\tiny$^\sim$}TMAD}), comprising data
from different sources, namely \citet{1975aelg.book.....B}, the
databases of the National Institute of Standards and Technology
(NIST\footnote{\url{http://www.nist.gov/pml/data/asd.cfm}}), the
Opacity
Project\footnote{\url{http://cdsweb.u-strasbg.fr/topbase/topbase.html}}
\citep[OP,][]{1994MNRAS.266..805S},
CHIANTI\footnote{\url{http://www.chiantidatabase.org}}
\citep{1997A&AS..125..149D,2013ApJ...763...86L}, as well as the
Kentucky Atomic Line
List\footnote{\url{http://www.pa.uky.edu/~peter/atomic}}. 

For calcium and the iron-group elements (Sc--Ni), we used a statistical approach
employing typically seven superlevels per ion linked by superlines,
together with an opacity sampling method
\citep{1989ApJ...339..558A,2003ASPC..288..103R}. 
Ionization stages {\tiny IV--VIII} augmented by a single, ground-level
stage {\tiny IX} were considered per species. We used the complete
line list of Kurucz \citep[so-called LIN lists, comprising about
$3.3\times10^7$ lines of the considered
ions;][]{kurucz1991,kurucz2009, kurucz2011} for the computation of
the non-LTE population numbers, and the so-called POS lists (that include only
the subset of lines with precisely known, experimentally observed line
positions) for the final spectrum synthesis.

For four trans-Fe elements (Zn, Ga, Ge, and Ba), we used the same
statistical approach. The atomic data and model atoms are taken from
our recent work on these species and are referred to below.

\section{Results}
\label{sect:results}

During our spectral-line fitting procedure, we found no indication that
the effective temperatures of both program stars
\citep{1998A&A...334..618D} require revision. That follows, for
instance, from the ionization balance of oxygen. We can evaluate lines
from three ionization stages in \pgvier\ and even four in
\pgsieben. We estimate that the temperatures are known to within $\pm
5000$\,K. As to the surface gravity, the \ion{He}{ii} lines fit well
with the gravity values derived by \citet{1998A&A...334..618D}. The
uncertainty in \logg\ is 0.5 dex.  In the following, we describe  the derivation of element abundances in
detail. For all comparisons of
model spectra to the FUSE observations in the shown figures, the
observations are smoothed with a 0.05\,\AA\ boxcar and 
the model spectra are convolved with a 0.1\,\AA\ (FWHM) Gaussian.

\subsection{Abundances of He, C, N, O}

The far-UV spectra of PG\,1159 stars are characterized by broad lines
that stem from \ion{He}{ii} and \ion{C}{iv} (and occasionally
\ion{N}{v}). Their profiles are shaped by linear Stark effect. In the
two stars considered here, the \ion{He}{ii} $n=2\rightarrow n'$ line
series is visible from $n'=5$ at 1085\,\AA, up to $n'=10$ at 949\,\AA,
with rapidly decreasing strength. The strongest \ion{C}{iii} lines at
977\,\AA\ and 1175--1176\,\AA\ are visible in both stars. As
\pgsieben\ is cooler, it also displays a number of more, weaker
\ion{C}{iii} lines. Table~\ref{tab:lines_c} lists all identified
carbon lines. 

As already mentioned, the first analysis of our two program stars
was based on medium-resolution optical spectra that only exhibited
lines from \ion{He}{ii} and \ion{C}{iv}. Only rough estimates for the
O abundance from the \ion{O}{v}~1371\,\AA\ line in low-resolution IUE
spectra could be made. For both stars, He = 0.33, C = 0.50, O = 0.17
was claimed \citep{1991A&A...244..437W}. Based on HST/GHRS UV spectra
(1165--1465\,\AA) taken with grating G140L (resolution 0.56\,\AA), which 
allowed for the detection of \ion{C}{iii-iv} and \ion{O}{iv-v} lines,
\citet{1998A&A...334..618D} slightly reduced the C abundance for
\pgsieben\ (He = 0.44, C = 0.39, O = 0.17) and the C and O abundances
for \pgvier\ (He = 0.50, C = 0.44, O = 0.06). From our investigation
of the FUSE spectra, the reduced C/He ratio in both stars is
confirmed, however, we verify the statement by
\citet{1998A&A...334..618D} that it can only be constrained within a
factor of two.

Because of the relatively high N abundance in \pgsieben, only this star
exhibits N lines. The presence of the \ion{N}{iv} multiplet at
1131--1133\,\AA\ was first noted by \citet{2007ASPC..372..237R}. We
identify an even larger number of \ion{N}{iv} lines. The strongest
\ion{N}{v} lines in the model are located at 1048--1050\,\AA\ (4d--4f
and 4f--6g transitions). It is blended by a similarly broad but
shallower \ion{C}{iv} feature of $n=4\rightarrow 11$ lines that are
not included in the model. Two \ion{N}{iii} lines are detected. One
near 979.9\,\AA\ is an unresolved multiplet and the other one, at
991.5\,\AA,\ are two unresolved lines of a resonance triplet. The
latter, however, is probably dominated by interstellar H$_2$.
Table~\,\ref{tab:lines_n} lists all identified nitrogen lines.

\begin{figure}[bth]
 \centering  \includegraphics[width=1.0\columnwidth]{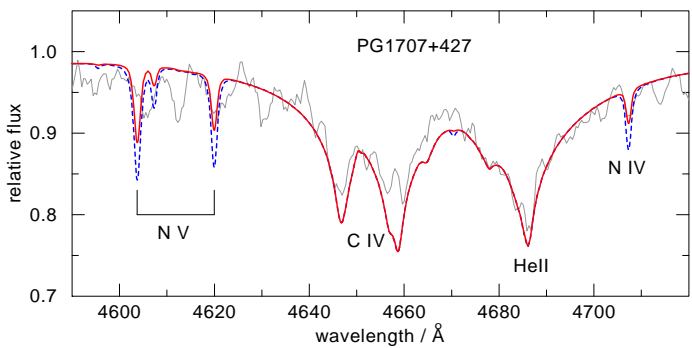}
  \caption{Section of the optical spectrum of \pgsieben\ { (thin
      line)} centered at the \ion{He}{ii}/\ion{C}{iv} absorption
    trough region. Overplotted are two models with different nitrogen
    abundance (N = 0.01 and 0.035{; thick solid and dashed lines,
      respectively}).  Other model parameters as listed in
    Tab.\,\ref{tab:stars}.}\label{fig:pg1707_opt}
\end{figure}

For their nitrogen abundance analysis, \citet{1998A&A...334..618D}
mainly relied on the \ion{N}{v} resonance doublet. They derived N =
0.035 for \pgsieben\ and an upper limit of N $< 3.5 \times 10^{-5}$
for \pgvier. While we cannot improve the limit for \pgvier, the N
abundance in \pgsieben\ requires a significant downward revision to N
= 0.01, according to our FUSE data analysis. The high value derived by
\citet{1998A&A...334..618D} must be regarded as an upper limit because
an interstellar contribution to the \ion{N}{v} resonance doublet
cannot be excluded. We  compared our new models with their GHRS
data and find that the observed \ion{N}{v} profile lies between the
two models with N = 0.035 and 0.0035, so that our low abundance from
the FUSE spectra is in reasonable agreement with the GHRS
data. Another constraint for the N abundance is the 3s--3p doublet,
located in the optical band at
4604/4620\,\AA. \citet{1998A&A...334..618D} considered their high N
abundance as compatible with the observation, however, their spectrum,
taken with the TWIN spectrograph at the Calar Alto 3.5m telescope, had
poor S/N. We have reobserved the star with the same equipment but
higher dispersion (36\,\AA/mm instead of 72\,\AA/mm) in 2001 and
display in Fig.\,\ref{fig:pg1707_opt} a detail of the spectrum around
the region of the absorption trough formed by \ion{He}{ii} and
\ion{C}{iv}. Overplotted are two model spectra with different N
abundance. It can be seen that the high abundance model overpredicts
the strength of the {N}{v} 4602/4620\,\AA\ doublet and, in addition,
that of a \ion{N}{iv} line at 4707\,\AA. The lower abundance fits
better. The observation is smoothed with a 0.8\,\AA\ boxcar and the
models with a 1\,\AA\ Gaussian. We have also inspected three optical
spectra of \pgsieben\ observed by the Sloan Digital Sky Survey,
however, they are of inferior quality.

As mentioned above, the far-UV spectra of the hot PG\,1159 stars
show many \ion{O}{vi} lines \citep[see Table~1
  in][]{2007A&A...462..281J}, which are as strong and broad as the
\ion{C}{iv} lines because of linear Stark broadening. No oxygen lines
of lower ionization stages are detectable in them. In the case of our
cool PG\,1159 stars, the situation is significantly different. Only a
small subset of \ion{O}{vi} lines is visible, in particular, the
resonance doublet in both of our stars plus a few subordinate lines in
\pgvier. In contrast, many lines from \ion{O}{iv} and \ion{O}{v} are
present. In fact, the majority of all identified far-UV lines in the
cool PG\,1159 stars originates from these two ionization stages. In
\pgsieben, we even identify a triplet of
\ion{O}{iii}. Table~\ref{tab:lines_o} lists all identified oxygen
lines.

\begin{figure}[bth]
 \centering  \includegraphics[width=1.0\columnwidth]{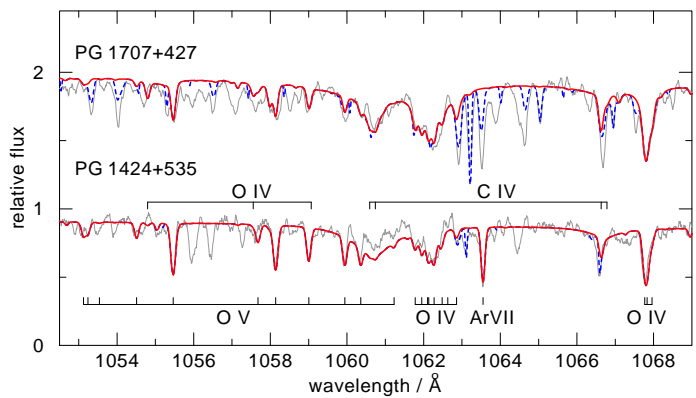}
  \caption{Detail of the FUSE spectra shown in Figs.\,\ref{fig:pg1707_fuse} and 
\ref{fig:pg1424_fuse} indicating many \ion{O}{iv} and \ion{O}{v} lines.
Interstellar lines in the models are plotted as {  dashed}.}\label{fig:oxygen}
\end{figure}

The oxygen abundance also needs a downward revision, namely to O =
0.03 for both stars, which is significantly lower in particular for
\pgsieben, so that the absolute abundances for He and C are scaled up
accordingly. In Fig.\,\ref{fig:oxygen} we show a particular wavelength
region in the FUSE spectra, where a number of prominent lines of
\ion{O}{iv} and \ion{O}{v} are located. The temperature difference
between both stars is reflected by the relative line strengths of both
ionization stages. Also, note the good fit to the \ion{O}{vi} resonance
doublet (second panels from top in Figs.\,\ref{fig:pg1707_fuse} and
\ref{fig:pg1424_fuse}), and the \ion{O}{iii} triplet in
\pgsieben\ (bottom panel of Fig.\,\ref{fig:pg1707_fuse}). 

\begin{figure}[bth]
 \centering  \includegraphics[width=1.0\columnwidth]{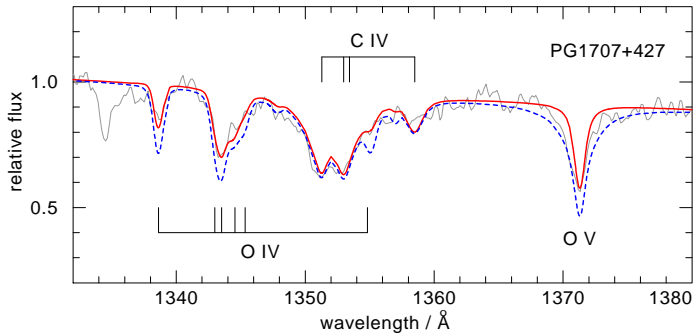}
  \caption{Section of the HST/GHRS spectrum of \pgsieben\ { (thin
      line)}. Overplotted are two models with different oxygen
    abundance (0.03 and 0.17{; thick solid and dashed lines,
      respectively}).  Other model parameters as listed in
    Tab.\,\ref{tab:stars}. }\label{fig:pg1707_ghrs}
\end{figure}

The low O abundance does not contradict the HST/GHRS data, which were
employed by \citet{1998A&A...334..618D}. In
Fig.\,\ref{fig:pg1707_ghrs} we show a detail of that observation
compared to two models that have the high abundance claimed by
\citet{1998A&A...334..618D} and the low abundance suggested by our
FUSE data analysis. In can be seen that the strengths of the
\ion{O}{iv} lines, as well as the \ion{O}{v}~1371\,\AA\ line, are
significantly overpredicted by the more O-rich model. The
GHRS observation has a resolution of 0.65\,\AA\ and was
smoothed with a 0.3\,\AA\ Gaussian. Accordingly, the model spectra
were smoothed with 0.72\,\AA\ Gaussians.

\subsection{Abundances of other metals}

Abundances of F, Ne, and Ar for both stars, and Fe in \pgvier\ were
determined in previous work, but is reconsidered here with our
new models. Abundances of Si, P, and S, as well as upper limits for
Ni, reported in our previous progress reports, are also
reassessed. Special care was taken to account for possible line
blends of the CNO elements. For the first time, we also checked for
the presence of Na, Mg, Al, and Ca, as well as the other iron-group
elements (Sc, Ti, V, Cr, Mn, Co) and selected trans-Fe group elements
(Zn, Ga, Ge, Kr, Xe, Ba). The typical error in the derived abundances
is 0.3~dex unless otherwise noted.

\subsubsection{Light metals (F--Ca)}

\paragraph{Fluorine.}

Three \ion{F}{v} lines (1082.3, 1087.8, 1088.4\,\AA) were detected in
both program stars, and \ion{F}{vi} 1139.5\,\AA\ in \pgvier. F =
10$^{-4}$ was derived for both objects \citep{2005A&A...433..641W}.
In \pgvier, the \ion{F}{vi} line at 1139\,\AA\ is best suited for a
model fit. In addition, we can use the two strongest components of the
\ion{F}{v} triplet at 1082\,\AA\ and 1088\,\AA. Care must be taken for
the latter because it is blended with an \ion{O}{v} line of
similar strength. In the present work, we find F = $5 \times 10^{-5}$
for \pgvier. The situation is more problematic for \pgsieben\ because
of the lower S/N of the spectrum and because \ion{F}{vi} 1139\,\AA\ is
not visible because of the lower temperature. The \ion{F}{v}
1088\,\AA\ line indicates F = $1.0 \times 10^{-4}$. Both results
essentially confirm our earlier analysis.

\paragraph{Neon.}

\ion{Ne}{vii} $\lambda\,973.33$\,\AA\ was discovered in \pgvier\ and
Ne = 0.02 was estimated \citep{2004A&A...427..685W}. The line is not
seen in \pgsieben\ because of its lower \Teff\ so that we cannot
assess the neon abundance in this star. The weak feature in the
model near this position is \ion{O}{iv}
973.43\,\AA\ (Fig.\,\ref{fig:pg1707_fuse}). It is also visible in the
model for \pgvier, hence, it blends the observed \ion{Ne}{vii}
line. This was unknown in our previous
analyses. For \pgvier, we derive Ne = $1.0 \times 10^{-2}$.

\paragraph{Sodium.}

We have performed detailed line formation calculations for
\ion{Na}{iv-vii}.  For both stars, there are no detectable lines in the
FUSE range of our spectral models even when the abundance is as large
as $10^{-2}$. No meaningful upper limits can be deduced.

\paragraph{Magnesium.}

In the models for both stars, two \ion{Mg}{iv} lines at 1044.36 and
1055.75\,\AA\ of a $^2$P--$^2$P$^{\rm o}$ multiplet can be used to
give an upper abundance limit (a third component with even larger $gf$
value is blended by the \ion{O}{vi} resonance line). We find Mg $\leq
5 \times 10^{-3}$.

\paragraph{Aluminium.}

In the model for \pgvier, the two strongest lines are from
\ion{Al}{iv}, namely, the 1125.61\,\AA\ component of a
$^3$S--$^3$P$^{\rm o}$ multiplet and a  $^1$P$^{\rm o}$--$^1$S singlet
at 1118.83\,\AA. They are used to set an upper limit of Al $< 5 \times
10^{-4}$. Weaker lines from \ion{Al}{v} are also exhibited by the
model, namely, the 1067.88\,\AA\ component of a $^4$P$^{\rm o}$--$^4$D
multiplet and the 1122.87\,\AA\ component of a $^2$D$^{\rm o}$--$^2$F
multiplet.  In the model spectrum of the cooler star, the \ion{Al}{iv}
lines are weaker and the \ion{Al}{v} lines have disappeared
completely. We find Al $< 5 \times 10^{-3}$.

\begin{figure}[bth]
 \centering  \includegraphics[width=1.0\columnwidth]{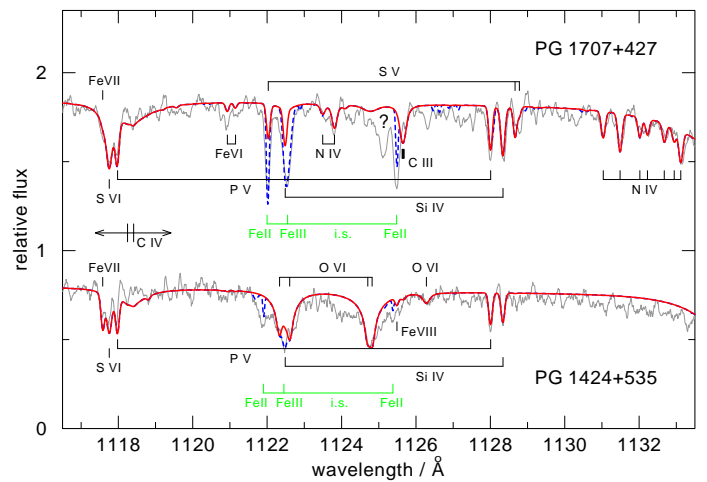}
  \caption{Detail of the spectra shown in Figs.\,\ref{fig:pg1707_fuse} and 
\ref{fig:pg1424_fuse} in a region with lines from Si, P, and S.
Interstellar lines in the models are plotted { as dashed}.}\label{fig:phosphorus}
\end{figure}

For the following discussion, we show  the spectral region
1117--1133\,\AA\ in detail in Fig.\,\ref{fig:phosphorus} to demonstrate a variety of the
discussed metal lines and the problem that some of them are blended
with interstellar lines. Most prominent in \pgvier\ are the broad
\ion{O}{vi} lines, which are essentially absent in \pgsieben\ because
of its lower temperature. 

\paragraph{Silicon.}

\citet{2005ASPC..334..225R} first identified Si in a PG\,1159 star by the
\ion{Si}{iv} 1122.49/1128.34\,\AA\ doublet in \pgvier. From that,
\citet{2007ASPC..372..237R} estimated Si = $3.6\times10^{-4}$. For
\pgsieben,\ they found Si = $7.2\times10^{-4}$. The presence of
\ion{Si}{v} lines, discovered by \citet{2007A&A...462..281J} in the
(hotter) prototype PG\,1159 star, is uncertain in \pgvier. The
\ion{Si}{iv} doublet and \ion{Si}{iv} 1066.63\,\AA\ are strong
diagnostic lines, however, the blue component of the doublet is
dominated by interstellar \ion{Fe}{iii} 1122.52\,\AA, and \ion{Si}{iv}
1066.63\,\AA\ is blended with a photospheric \ion{C}{iv} line at the
same wavelength and interstellar \ion{Ar}{i} 1066.66\,\AA. Hence,
relying on the 1128.34\,\AA\ line alone, we find Si = $3.2 \times
10^{-4}$ for \pgsieben\ and $2.0 \times 10^{-4}$ for \pgvier, which are
about half the values found previously.

\paragraph{Phosphorus.}

The detection of P in a PG\,1159 star was first reported by
\citet{2005ASPC..334..225R}, who found the \ion{P}{v}
1118/1128\,\AA\ resonance doublet in \pgvier.
\citet{2007ASPC..372..237R} estimated P = $9.5\times10^{-6}$ and
$2.0\times10^{-6}$ for \pgvier\ and \pgsieben, respectively. No other
P lines are detectable. For the abundance analysis, care must be taken
in the case of \pgsieben\ because the 1117.98\,\AA\ component of
\ion{P}{v} is coinciding with a \ion{N}{iv} line at 1117.93\,\AA\ that
is slightly weaker. We find P = $1.0 \times 10^{-5}$. This is a
compromise because at that abundance the 1118\,\AA\ component in the
model is slightly too strong, while the 1128\,\AA\ component  is too
weak compared to the observation. Since \pgvier\ has no detectable
nitrogen, the \ion{P}{v} 1118\,\AA\ component can safely be used for
the abundance determination. Note that the line is located in the blue
wing of a \ion{C}{iv} line, and right between two lines of \ion{S}{vi}
and \ion{Fe}{vii}. We find P = $3.2 \times 10^{-5}$. 

\paragraph{Sulphur.}

The presence of the \ion{S}{vi} 933/945\,\AA\ resonance doublet in
both stars was first noted by \citet{2005ASPC..334..225R}. They also
reported the possible identification of \ion{S}{vi} 1117.76\,\AA\ in
\pgvier. From the resonance doublet, \citet{2007ASPC..372..237R}
derived S = $5.0\times10^{-5}$ and $2.5\times10^{-5}$ for \pgvier\ and
\pgsieben, respectively. Another \ion{S}{vi} doublet at
1000.5\,\AA\ is detectable in both stars, and a few \ion{S}{v} lines
are seen in \pgsieben\ (Table~\ref{tab:lines}). For \pgsieben,\ the
best fit is achieved at S = $3.2 \times 10^{-4}$. \ion{The S}{v} line at
1122\,\AA\ is dominated by an interstellar
\ion{Fe}{ii} line. In \pgvier, no \ion{S}{v} line is seen
because of its higher temperature. From the \ion{S}{vi} lines, we find
S = $1.0 \times 10^{-4}$. For both stars, the derived abundances are
significantly higher than the previous estimates. 

\paragraph{Argon.}

The hitherto only detection of Ar in a PG\,1159 star is the
\ion{Ar}{vii} 1063\,\AA\ line in
\pgvier\ \citep{2007A&A...466..317W}. Ar = $3.2 \times 10^{-5}$ was
derived, however, our reanalysis yields about twice that value,
namely, Ar = $6.0 \times 10^{-5}$, which is close to the solar
value. The line is not accessible in \pgsieben\ because it is blended
by interstellar H$_2$.

\paragraph{Calcium.}

There are no Kurucz POS \ion{Ca}{v-vii} lines in the FUSE range. Lines
from other Ca ions are not detectable in the models. No abundance
constraints for calcium can be found.

\subsubsection{Iron-group elements (Sc--Ni)}

The derivation of the iron abundance and at least upper limits for the
other iron group elements is of interest because they allow us to
constrain the effects of radiative levitation. For example, extremely
large overabundances were observed in hot subdwarfs due to this
process \citep{2009ARA&A..47..211H}.

For scandium, only lines from ionization stage \ion{Sc}{v} are
available in the used line lists. They have relatively weak oscillator
strengths. No meaningful upper abundance limits can be derived
because even at $10^4$ times solar Sc abundance no significant line is
detected in the model spectra. For titanium, only lines from
\ion{Ti}{v} are in the line lists for the FUSE range. They are very
weak in the models for \pgsieben\ and even weaker in the models for
\pgvier. An upper limit can be derived only for the former, from the
\ion{Ti}{v} line blend at 1153.26 and 1153.28\,\AA.  A few very weak
vanadium lines appear in the models. From the three strongest,
\ion{V}{v} 1142.74, 1157.57, and 1159.52\,\AA, we derive upper
abundance limits. In the models for \pgsieben, the strongest chromium
lines are \ion{Cr}{v} 1104.30\,\AA\ and \ion{Cr}{vi} 957.01\,\AA. For
\pgvier, the \ion{Cr}{v} line disappears and the \ion{Cr}{vi} line
becomes weaker. Only weak manganese lines, \ion{Mn}{v} (1055.98\,\AA)
and \ion{Mn}{vi} 927.61, 933.78, 1113.58\,\AA, are noticeable in the
models. 

\ion{Fe}{vii} lines were detected in \pgvier\ and a solar abundance
was derived \citep{2011A&A...531A.146W}. Two \ion{Fe}{viii} lines are
possibly present in \pgvier\ (Table~\ref{tab:lines}); they were first
discovered in hotter PG\,1159 stars by \citet{2011A&A...531A.146W}.
Because of the lower temperature, the \ion{Fe}{vii} lines are not
clearly detectable in \pgsieben. Weak lines from
\ion{Fe}{vi} could be expected but their identification is uncertain as well (Table~\ref{tab:lines}). According to our models, the strongest
line of \ion{Fe}{v} is at 1005.96\,\AA, but it is blended with
\ion{N}{iii}. We confirm the solar abundance in \pgvier\ found
previously. Since iron lines cannot be detected beyond doubt, we can
only derive an upper limit for \pgsieben. 

Only rather weak cobalt lines of \ion{Co}{vi} (e.g., 1128.75 and
1172.88\,\AA) are present in the models. Our models with solar Ni
abundance predict only weak nickel lines. The four strongest are from
\ion{Ni}{vi} (Table~\ref{tab:lines}). In the model for \pgsieben,
their strengths are just below the detection limit regarding the S/N
of the observation. An upper abundance limit of solar can be derived,
confirming \citet{2008ASPC..391..121R}. Because of the higher \Teff,
the \ion{Ni}{vi} lines in the model for \pgvier\ are even weaker. No
nickel line of higher ionization stages is available in the Kurucz POS
lists. 

\begin{figure*}[bth]
 \centering  \includegraphics[width=0.95\textwidth]{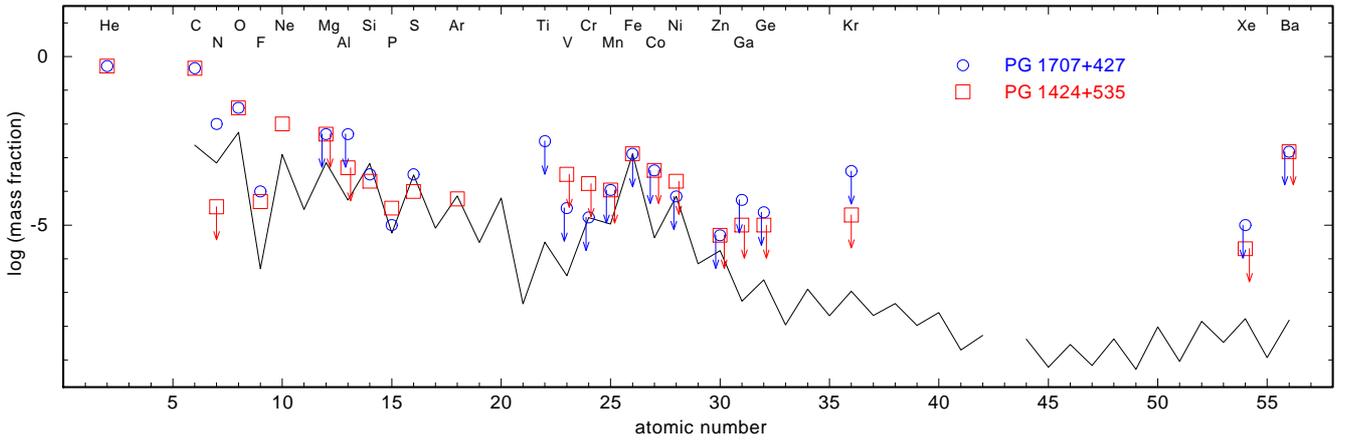}
  \caption{Derived elemental abundances. Upper limits are marked with
    arrows. Solar abundance values are indicated with the black
    line.}\label{fig:abundances}
\end{figure*}

\subsubsection{Trans-iron group elements: Zn, Ga, Ge, Kr, Xe, and Ba}

These elements have been detected in the FUSE spectrum of the hot DO
white dwarf RE\,0503$-$289 (\Teff = 70\,000\,K, \logg = 7.5) with
extreme overabundances. The star is considered a PG\,1159--DO
transition object and it is therefore interesting to check whether these
elements are overabundant in PG\,1159 stars as well. In particular, six
species were studied in detail, namely, Kr and Xe
\citep{2012ApJ...753L...7W}, Zn \citep{2014A&A...564A..41R}, Ga
\citep{2015A&A...577A...6R}, Ge \citep{2012A&A...546A..55R}, and Ba
\citep{2014A&A...566A..10R}. In those papers, the strongest lines of
these elements were identified in the models and the observation. None
of these species is detected in our stars and we derived the upper
abundance limits listed in Tab.\,\ref{tab:stars} after the following
considerations.

The models for both stars exhibit a large number of \ion{Zn}{v}
lines. The strongest are located at 1116.85, 1138.25, and
1142.93\,\AA, and there is a strong blend of several lines at
1177\,\AA. Numerous lines of \ion{Ga}{v} and \ion{Ga}{vi} are seen in
the models for \pgsieben, while in the models for \pgvier, only
\ion{Ga}{vi} lines are present. Just as many lines of \ion{Ge}{v}
and \ion{Ge}{vi} in the models behave like the respective gallium
lines. The strongest features are \ion{Ge}{v} 1045.71 and
1072.66\,\AA\ as well as \ion{Ge}{vi} 926.82\,\AA. The strongest
krypton line in the models is \ion{Kr}{vii} 918.14\,\AA, and the
strongest xenon lines are \ion{Xe}{vii} 995.51 and 1077.12\,\AA.  In
the models for \pgsieben, the strongest lines of barium are from
\ion{Ba}{vi} (937.24 and 953.39\,\AA) and \ion{Ba}{vii} (943.10 and
993.41\,\AA). In the models for \pgvier, the \ion{Ba}{vi} lines are no
longer visible.

\section{Summary and discussion}
\label{sect:discussion}

We have analyzed the far-UV spectra of two relatively cool PG\,1159
stars  to derive elemental abundances and to compare them with
predictions from stellar-evolution theory. The only other PG\,1159
star hitherto investigated in similar detail is the prototype itself
\citep[\elf,][]{2007A&A...462..281J}. The resulting abundance values
are listed in Table~\ref{tab:stars} and displayed in
Fig.\,\ref{fig:abundances}.  The analysis was performed with non-LTE
model atmosphere calculations employing vastly extended and new model
atoms as compared to our earlier, exploratory assessments of these
stars. Complete compilations of all identified photospheric lines are
presented. 

\begin{figure}[bth]
 \centering  \includegraphics[width=1.0\columnwidth]{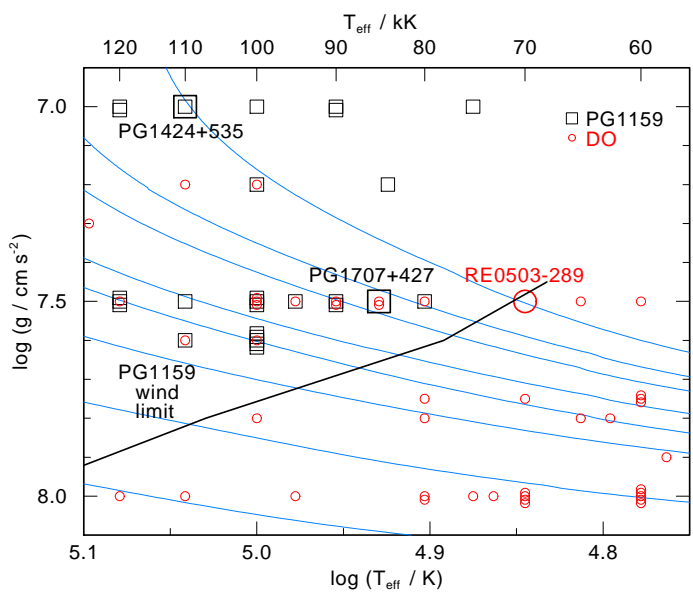}
  \caption{Position of PG\,1159 stars (squares) and DO white dwarfs
    (circles) in the \logg--\Teff diagram. Our two program stars and
    a particular DO are indicated with big symbols and name
    tags. Blue lines are evolutionary tracks by
    \citet{2009ApJ...704.1605A} with stellar masses of 0.514, 0.530,
    0.542, 0.565, 0.584, 0.609, 0.664, and 0.741 \Msol\ (from top to
    bottom). The black line is the PG\,1159 wind limit (see text).}\label{fig:gteff}
\end{figure}

Considering the numerous ionization balances of the different species,
we found no hint that effective temperatures and surface gravities
determined by previous work based on optical and UV spectra require
revision. Consequently, the masses derived by comparison with
evolutionary tracks
\citep[Fig.\,\ref{fig:gteff},][]{2006PASP..118..183W,2009ApJ...704.1605A}
also remain unaltered. The masses are M = 0.53 and 0.51~\Msol\ for
\pgsieben\ and \pgvier, respectively. Considering the initial-final
mass relation by \citet{2000A&A...363..647W}, this corresponds to
about one solar mass for their main-sequence progenitors. However, the
error in \logg (0.5 dex) means that the masses could be as high as 0.7
and 0.58~\Msol, corresponding to 3.5 and 2.5~\Msol\ progenitors. 

While we confirmed the C/He ratio in the stars, the mass fraction of
oxygen was found to be significantly lower compared to earlier work
\citep{1998A&A...334..618D}. For both stars, we find O = 0.03 instead
of 0.17 and 0.07 for \pgsieben\ and \pgvier, respectively. This value
is relatively low compared to the prototype \elf\ with O = 0.17, but
identical to three other cool PG\,1159 stars reanalyzed recently with
high-resolution optical spectra \citep{2014A&A...569A..99W}. Likewise,
the N abundance in \pgsieben\ was reduced from N = 0.035 to 0.01, and
the Ne abundance in \pgvier\ from Ne = 0.02 to 0.01. Concerning trace
elements, the abundances of other light metals (F, Si, P, S, Ar) were
reassessed and, in some cases, revised. The solar Fe abundance in
\pgvier\ was confirmed. Upper limits for most other iron-group
elements (Ti, V, Cr, Mn, Co, Ni) as well as some trans-iron elements
(Zn, Ga, Ge, Kr, Ce, Ba) were derived for the first time.

Our discussion on the element abundances is mainly based on a
comparison with stellar evolution models by
\citet{2013MNRAS.431.2861S}, who presented intershell abundances for
stars with initial masses of 1.8--6~\Msol. They compared their results
to abundances derived for the prototype \elf.  In general, good
agreement was found for the species investigated (He, C, O, F, Ne, Si,
P, S, Fe) with the notable exception of sulphur, for which a strong
depletion (0.02 times solar) in \elf\ was claimed by
\citet{2007A&A...462..281J}, who speculated that the very slight
depletion of S by n-capture nucleosynthesis predicted by stellar
models (0.6 solar) could point to modeling problems of the AGB
intershell chemistry. In addition, one possible solution of the
so-called sulphur anomaly in planetary nebulae, which describes an
unexplained S depletion \citep{2004AJ....127.2284H}, could originate
from sulfur transformation into chlorine by neutron captures and
subsequent beta decay in the central star's preceding AGB
phase. Consequently, \citet{2013MNRAS.431.2861S} have examined to
what extent particular uncertainties of nucleosynthesis modeling
(nuclear reaction rates, mass width of partially mixed zone) can
affect the resulting intershell abundances of stars with a variety of
initial masses. However, no explanation for the sulphur anomaly could
be found.  In contrast to the prototype, in the two stars of the
present study we find S = 1/3 solar and solar, which is in agreement
with stellar models. 

While the carbon abundance in our two stars (C = 0.45) is in agreement
with the value found for the prototype (C = 0.48), the oxygen
abundance is significantly lower (O = 0.03 compared to
0.17). Nevertheless, the various C and O abundances are consistent
with models that include overshoot into the C--O core
\citep{2006PASP..118..183W}. (Such high values are not exhibited in
the \citet{2013MNRAS.431.2861S} models because they do not include
that strong overshoot.) As to nitrogen, the presence of ample N in one
of our stars was already discovered by \citet{1998A&A...334..618D}. It
is explained by the fact that the late thermal pulse had occurred in
this star when it was already a WD (so-called very late TP), leading to
$^{14}$N production by H ingestion and burning. The absence (or much
smaller) N abundance in the other star is the consequence of the late
thermal pulse that the progenitor suffered during its previous H-shell
burning post-AGB phase.

Our study has confirmed the previously found extreme overabundance of
fluorine in our two stars as well as other PG\,1159 stars including
the prototype, which is consistent with stellar models \citep[see][and
  discussion therein]{2005A&A...433..641W}. In particular, the
abundances found in \pgsieben\ and \pgvier\ (10 and $5\times10^{-5}$)
are consistent with the range of abundances in models with different
mass and reaction rates presented by \citet{2013MNRAS.431.2861S} (2--8
$\times10^{-5}$), although their lowest mass model (1.8~\Msol) falls
short by factors of 2.5 and 5 to correctly predict the observed high F
abundances. 

Generally, neon was found to be strongly enriched  in PG\,1159 stars
to typical values of Ne = 0.02 \citep{2004A&A...427..685W}, in
agreement with stellar models (where $^{14}$N is transformed into
$^{22}$Ne by two $\alpha$ captures). The neon abundance we find in
\pgvier\ is half that value but, regarding our error estimate, this is
not significantly below the predictions.

Our results on the silicon abundance (0.3--0.5 solar) are, within
error limits, consistent with the model predictions by
\citet{2013MNRAS.431.2861S} (0.7--0.9 solar). For phosphorus, we find
5.5 solar for \pgvier, matching the P overproduction of four times
solar predicted by the 1.8~\Msol\ model of
\citet{2013MNRAS.431.2861S}. For \pgsieben, 1.7 solar was determined,
matching the 1.9 solar overproduction of P in a 3.0~\Msol\ model.

We have investigated the possibility of discovering other light metals
that were not detected before in PG\,1159 stars. It turned out that
Na, Mg, Al, and Ca are undetectable even at strongly oversolar
abundances (Tab.\,\ref{tab:stars}).

The solar iron abundance found in \pgvier\ \citep[and other PG\,1159
  stars,][]{2011A&A...531A.146W} was confirmed and is in agreement
with the \citet{2013MNRAS.431.2861S} stellar models. They predict a
slight Fe reduction of about 30\%, a small amount that is not
verifiable by our analysis. For all other iron-group elements except
of scandium, upper abundance limits were derived. Their nondetection
is consistent with solar values and excludes overabundances of more
than 1--2 dex as detected in hot subdwarfs as a consequence of
radiative levitation \citep{2009ARA&A..47..211H}.

For the first time, we  assessed the question as to the presence
of trans-Fe elements in PG\,1159 stars. This was inspired by the
recent discovery of extreme overabundances in the hot DO white dwarf
\re\ (\Teff = 70\,000\,K, \logg = 7.5, M = 0.51~\Msol), which is
believed to be an object  in the PG\,1159--DO transition phase
(Fig.\,\ref{fig:gteff}) evolving across the so-called wind limit
\citep{2000A&A...359.1042U}, when gravitational settling and radiative
acceleration of elements begin to affect the chemical surface
composition.  For six elements (Zn, Ga, Ge, Kr, Xe, Ba), 155--23\,000
times solar values were derived \citep[][and references
  therein]{2015A&A...577A...6R}, corresponding to mass fractions in
the range 0.3--3.5 $\times10^{-4}$. We have not found
any trace of these species in the two investigated PG\,1159 stars, but
from the determined upper abundance limits (Table\,\ref{tab:stars})
the following can be concluded. 

We do not see an enrichment of these elements as extreme as in \re, in
particular, the abundances of Zn, Ge, and Xe in \re\ are above the
detection thresholds in \pgsieben\ by factors of 54, 6, and 6, 
 and by factors of 54, 15, 31 in \pgvier, respectively. In the latter
star,  Ga and Kr, if present as abundant as in \re, would also be above
the detection threshold by factors of 2 and 3. It is probable that in
\re\ these elements are driven to such very high abundances by
radiative levitation.

What abundances of the trans-iron elements do we find in the
intershell of AGB star models and how do they compare to our detection
thresholds?  In stellar AGB nucleosynthesis models of a 2\,\Msol\ star
\citep[Gallino priv. comm.,][]{2007ApJ...656L..73K}, the following
values are found after the 30th thermal pulse: Zn =
$2.2\times10^{-6}$, Ga = $2.7\times10^{-7}$, Ge = $5.9\times10^{-7}$,
Kr = $1.9\times10^{-6}$, Xe = $1.5\times10^{-6}$, and Ba =
$1.2\times10^{-5}$. A comparison with Table\,\ref{tab:stars} shows that
the detection thresholds of zinc in both stars (Zn =
$2.2\times10^{-6}$) and of xenon in \pgvier\ (Xe = $2.0\times10^{-6}$)
are just barely higher, while the abundances of the other elements are
well below the respective limits. Spectra of better S/N would allow
the detection of at least Zn, Xe, and Kr.

In summary, far-UV spectroscopy of relatively cool PG\,1159 stars is
essential to determine abundances of light and heavy metals beyond C,
N, and O. The light metals (F--Ar) are generally in good agreement
with evolution models. We do not find extreme overabundances of
iron-group and trans-iron group elements, which is  consistent with the
expectation that radiative levitation and gravitational settling are
not affecting the abundance patterns in these stars. The detection of
trans-iron elements would be valuable to constrain parameters of the
s-process nucleosynthesis modeling, but far-UV spectra with better S/N
are required.

Finally, a particular aspect of the pulsation properties of PG\,1159
stars is interesting.  Nonpulsating stars exist in the
respective GW~Vir instability strip. The nitrogen dichotomy mentioned
in Sect.\,\ref{sect:programstars} led \citet{1998A&A...334..618D} to
conclude that nitrogen is an essential ingredient for pulsation
driving because the nonpulsators have low N abundances. On the other
hand, it was shown that it is predominantly the high O abundance that
is responsible for pulsation driving
\citep[e.g.,][]{2007ApJS..171..219Q}. We found that our two
program stars have the same O (and C) abundance, while the N
abundance is significantly different. Curiously, \pgsieben\ (N = 0.01)
is a pulsator and \pgvier\ (N $< 3.5 \times 10^{-5}$) is a
nonpulsator, which would confirm the conclusion by
\citet{1998A&A...334..618D}. We  recently obtained a similar
result from an analysis of three other cool PG\,1159 stars
\citep{2014A&A...569A..99W}.

\begin{acknowledgements} 
T. Rauch is supported by the German Aerospace Center (DLR) under grant
05\,OR\,1402. This research has made use of the SIMBAD database,
operated at CDS, Strasbourg, France, and of NASA's Astrophysics Data
System Bibliographic Services. Some of the data presented in this
paper were obtained from the Mikulski Archive for Space Telescopes
(MAST). This work had been done using the profile fitting procedure
Owens.f, developed by M\@. Lemoine and the FUSE French Team. 
\end{acknowledgements}

\bibliographystyle{aa}  \bibliography{aa}

\begin{appendix}\label{sect:app}
\section{Model spectra including ISM lines}

\begin{figure*}[bth]
 \centering  \includegraphics[width=1.0\textwidth]{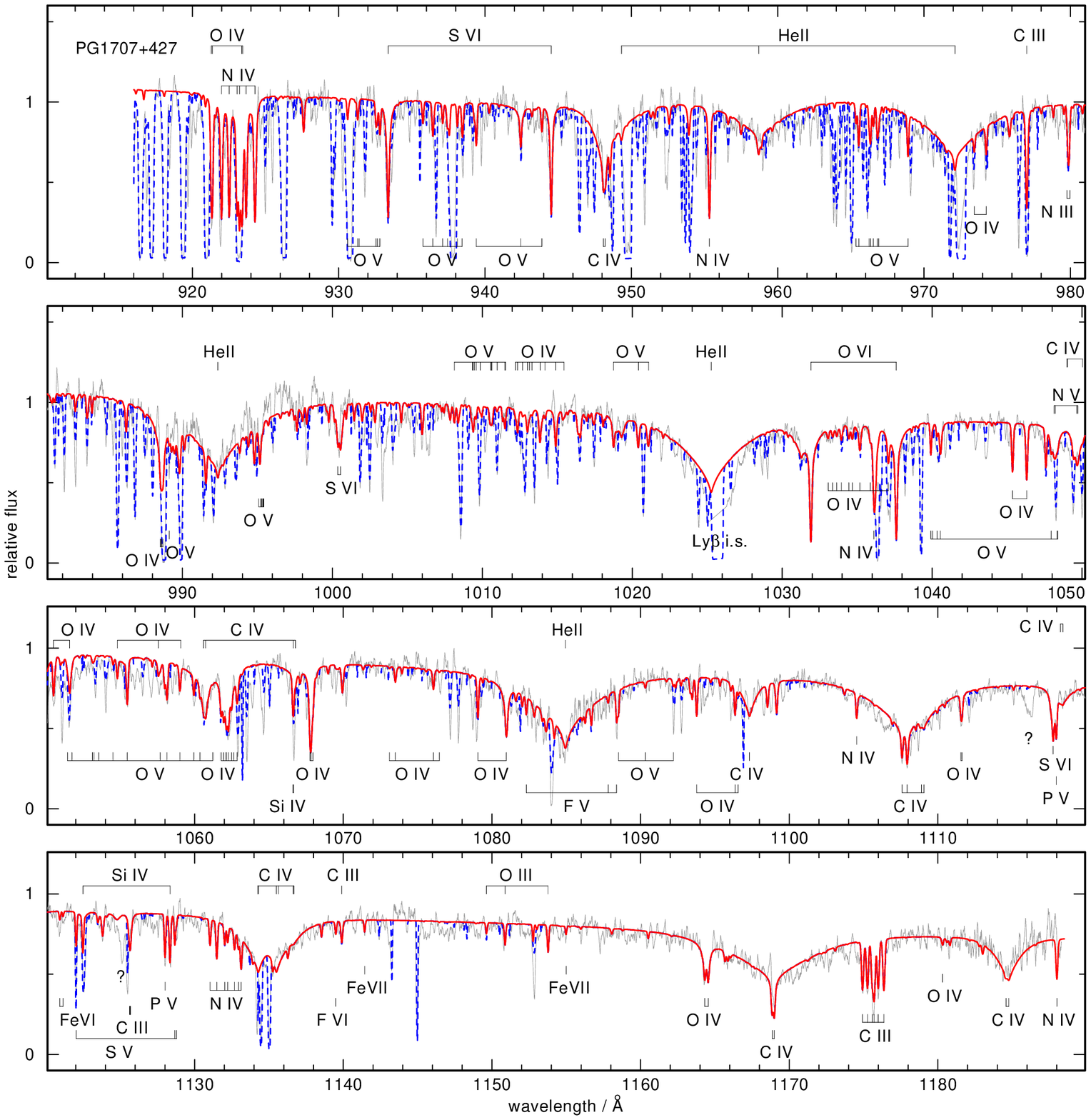}
  \caption{As in  Fig.\,\ref{fig:pg1707_fuse}, except that it includes ISM lines { (dashed graph)}.}\label{fig:pg1707_fuse_ism}
\end{figure*}

\begin{figure*}[bth]
 \centering  \includegraphics[width=1.0\textwidth]{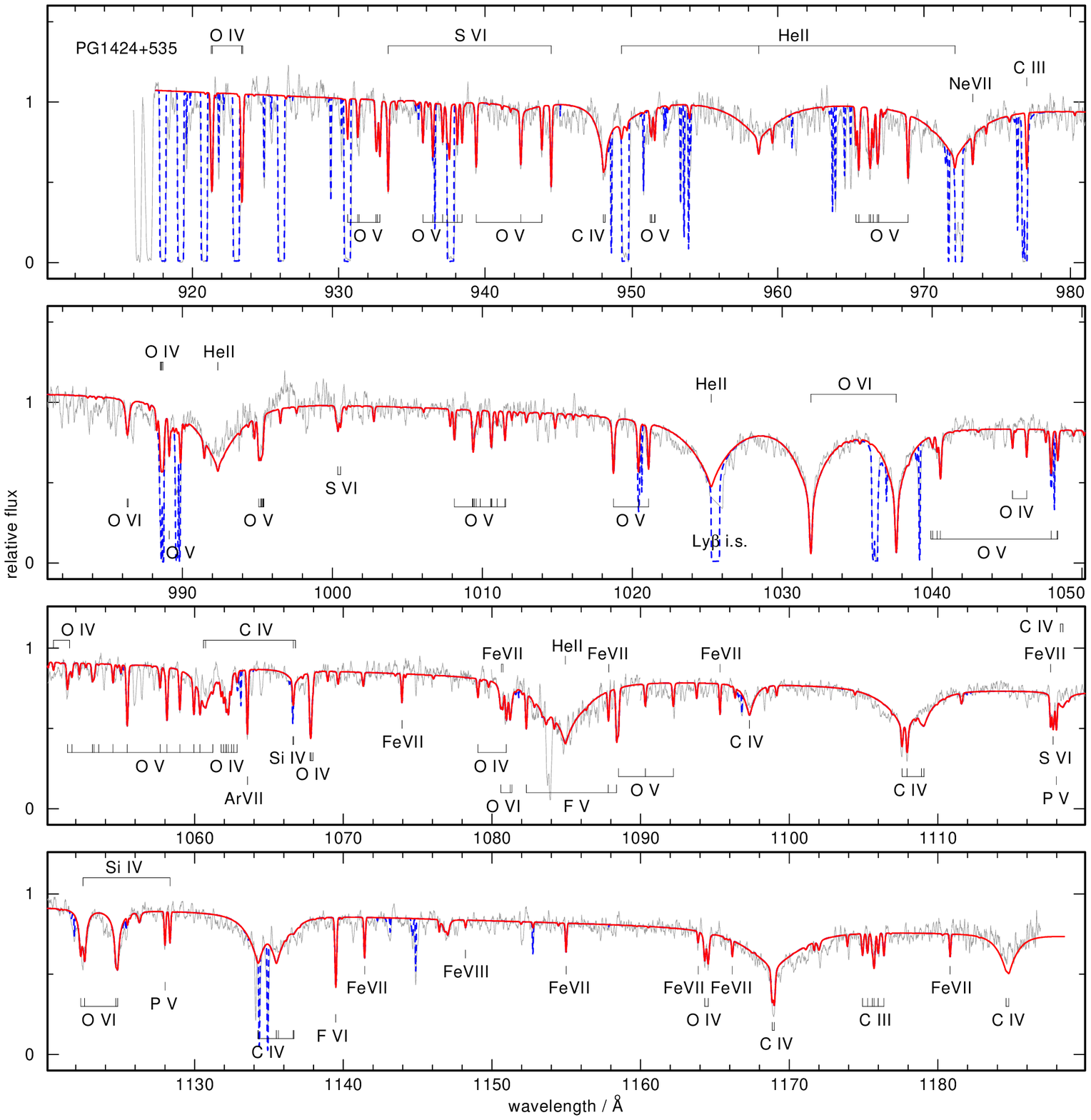}
  \caption{As in  Fig.\,\ref{fig:pg1424_fuse}, except that it includes ISM lines { (dashed graph)}.}\label{fig:pg1424_fuse_ism}
\end{figure*}

Figures \ref{fig:pg1707_fuse_ism} and \ref{fig:pg1424_fuse_ism}  are
identical to Figs.\,\ref{fig:pg1707_fuse} and \ref{fig:pg1424_fuse},
except they include the ISM lines. The pure photospheric spectrum is plotted
in red, while the additional ISM lines are indicated in blue.

\end{appendix}

\end{document}